\documentclass[preprint,aps,showpacs,prc,nofootinbib]{revtex4-1}
\usepackage{epsfig}

\newcommand{\LQ}{``}
\newcommand{\Be}{\begin{equation}}
\newcommand{\Ee}{\end{equation}}
\newcommand{\Bea}{\begin{eqnarray}}
\newcommand{\Eea}{\end{eqnarray}}

\begin{document}

\title{A Mesonic Analog of the Deuteron}
\author{T.\ Goldman}\email{tgoldman@lanl.gov}
\affiliation{Theoretical Division, MS-B283, 
	Los Alamos National Laboratory, Los Alamos, NM 87545 \\ 
	{\rm and} \\
	Dept. of Physics and Astronomy, University of New Mexico, 
	Albuquerque, NM 87501}
	\author{Richard R.\ Silbar}\email{silbar@lanl.gov}
\affiliation{Theoretical Division, MS-B283, \\
	Los Alamos National Laboratory, Los Alamos, NM 87545}

\begin{flushright}
December 20, 2013\\
{LA-UR-13-22745}\\
{arXiv:1304.5480}\\
\end{flushright}

\begin{abstract}
Using the LAMP model for nuclear quark structure, we calculate the 
binding energy and quark structure of a $B$ meson merging with a $D$ meson. 
The larger-than-nucleon masses of the two heavy quarks allow for a 
more reliable application of the Born-Oppenheimer-like approximation 
of the LAMP.  With the absence of quark-level Pauli Exclusion 
Principle repulsive effects, the appearance of a bound state is unsurprising.
Our variational calculation shows that the molecular, deuteron-like state
structure changes rather abruptly, as the separation between the two mesons 
decreases, at a separation of about 0.45 fm, into a four-quark bound state, 
although one maintaining an internal structure rather than that of a four-quark 
bag.  Unlike the deuteron, pion exchange does not provide any contribution 
to the $\approx 150$ MeV binding. 

\end{abstract} 

\maketitle

\section{Introduction}

What would nuclear physics look like without pion exchange? The long range 
of the nuclear force due to pion exchange between nucleons, along with the 
empirical short distance repulsion between nucleons, supports the established 
view of nuclear physics as due to the interaction of effective degrees of freedom 
that bear a very close resemblance to free space nucleons. Calculations of nuclear 
structure for small nuclei, using potential interactions fit to scattering data, succeed 
quite accurately.\cite{Carlson} Effective field theory expansions, with or without 
pions, claim successes~\cite{Bira} as well. For large nuclei, elaborations of the shell 
model can also reproduce experimentally known results. 

However, all of these approaches ignore the internal structure of the three-quark 
states that are on-shell nucleons in free space but not so well defined off-shell 
degrees of freedom in the nucleus. In particular, the basis for off-shell nucleon 
form factors resembling those of on-shell nucleons is weak, and conflicts with 
the experimental results of deep inelastic scattering (DIS) on nuclei. Those results 
are {\em not} well represented by multiplying the results of DIS on free space 
nucleons by the number of nucleons in the target nucleus. This is known as the 
\LQ EMC effect".\cite{EMC}

The relativistic Los Alamos Model Potential~\cite{GMSS,GBS} (LAMP) has been 
used to describe the binding and structure of $^3$He and $^4$He, including a 
good description~\cite{BG} of the deep inelastic structure function of $^3$He. 
It was explicitly constructed to access the internal quark structure of the baryonic 
components of the nucleus without the presumption of a free space nucleon 
approximation. As such, except for the difficulties of carrying out calculations, 
it provides a less biased view (although not a systematic expansion) 
of the hadronic structure of nuclei than do the conventional models referred to 
above. 

The LAMP does not describe the deuteron at all due to the very large separation 
of the nucleons and the dominance of single-pion exchange contributions there.\cite{Friar}
The LAMP, lacking quark-exchange correlations, best encompasses medium 
and short-range meson exchanges (two-pion, $\rho$, etc.). It must therefore be 
supplemented with long-range single-pion-exchange contributions \cite{piqk} 
for a better description of nuclear binding energies.  

However, in this model, we {\em can} ask: What would nuclear physics, and in particular, the deuteron, look 
like in the absence of long-range pion exchange interactions? If bound states exist, 
the constituents would be much closer together than in actual nuclei and disruption 
of the internal structure could be much more significant than suggested by the LAMP 
as applied to nucleons 
or the results of conventional nuclear physics. Could one still identify nucleonic 
effective degrees of freedom even when the multi-quark hadronic objects are in such 
close proximity that their average separation is less than their internal structure? This 
is to be contrasted with real nuclei where the mean separation between nucleons is 
quite close to twice their root-mean-square radii. 

In this paper, we make an initial address to this question by considering a simpler 
problem, the binding of two heavy mesons. Large mass quarks are used to mimic 
the large mass of the nucleon, but one light antiquark in each stands in for the diquark 
in the nucleons and so simplifies the calculations. Since no quark-exchange correlations 
are included, no ($t$-channel) quark-antiquark combinations with pion quantum numbers 
contribute any more significantly than higher mass mesons. However, the extension/size 
of the mesonic states is comparable to that of nucleons due to the spread of the light quark 
wavefunction. 

In fact, for this case, all \LQ light" meson exchanges are prevented, and the interactions 
have solely to do with the structure of the light antiquark wave functions under the influence 
of the color confining force, represented here by a collective potential. This is somewhat 
analogous, in principle, to the nuclear shell model potential although significantly different 
in form to be consistent with known models of confinement. 

In particular, we examine here the structure of a four-quark system derived from  
$B^{-} = b\bar{u}$ and $D^{+} = c\bar{d}$ mesons for a bound state, or their neutral 
equivalents when the light antiquarks are exchanged between them. Because these 
mesons are considerably more massive than nucleons, localization energy is much 
reduced.  This brings them into closer proximity than the nucleons in a deuteron, or indeed, 
even in a large nucleus. The larger-than-nucleon masses of the two heavy quarks 
also allow for a more reliable application of the Born-Oppenheimer-like approximation 
of the LAMP.  Furthermore, the quark content chosen here does not involve any pairs 
of quarks with the same (internal) quantum numbers, so there are no (quark) Pauli 
exclusion effects such as those that contribute to the short-range repulsion between 
nucleons. Thus, this is a system in which one can expect greater accuracy of the LAMP 
and a significantly more deeply bound state than the deuteron. 

When this $B$-$D$ bound state is observed, the deviation from our predictions here 
will provide a very good measure of the center of mass motion and breathing mode collective 
excitations. These are difficult to remove in the LAMP due to its relativistic nature. Since the 
non-relativistic model analogous to the LAMP, the Quark Delocalization and Color Screening 
Model of Wang {\it et al.\ }\cite{fanwang}, gives very similar results to the LAMP after removing 
such effects, we expect the corrections due to these effects to be small.
Thus our predictions here should be reasonably accurate. 

There have been many different approaches, going back to the Cornell potential \cite{eichten}, 
along lines comparable to the LAMP, to modelling quark-antiquark states using potentials. We 
note here a few recent references \cite{others}. There have also been many papers devoted to 
the study of four-quark systems, with a view to identifying exotic states constructed of more than 
three quarks or one quark and one antiquark. See, for example, the references in the recent 
review of Brambilla {\it et al.\ }\cite{nora} and some very early papers \cite{heller} as well.  
Generally, however, these papers have focused on states more likely to appear in hadronic 
collisions, such as those with the quark content of $B$ and $\bar{B}$ or $D$ and $\bar{D}$ 
mesons and their excited state partners (for a recent example, see \cite{PB}), since strong production 
of heavy quarks proceeds in a pairwise fashion. (Some, such as Ref.(\cite{heller}), have also included 
consideration of the case studied here, albeit without the intricacies available in the LAMP). 
In general, the mixing of these states with the charmonium and bottomonium spectra, however, make 
for difficulties in extracting  them unambiguously from experimental observations and may require the 
determination of exotic quantum numbers. No such problems occur in the case considered here, 
although the reduced probability of production must certainly be recognized. In any event, our interest 
is not in the prediction of exotic states, but in the elucidation of the origins of the nature of nuclear 
structure and thus a deeper understanding of it.

\subsection{Initial Concepts}

The LAMP treats the confining potential for quarks (and antiquarks) as a fixed scalar interaction 
in a Born-Oppenheimer-like picture, with the location of the potential minimum defining the 
system location. Quarks bound in a baryon or meson are treated as being bound within this 
potential rather than directly to each other. As such, there are immediate concerns about removing 
center-of-mass and breathing mode contributions to the evaluated state energy. This concern 
is ameliorated by comparing the energy of the interacting system of the two heavy mesons with 
the value at large (essentially infinite) separation. 

In this paper, in addition to the confining Lorentz scalar potential of the LAMP, we have included 
a Lorentz vector potential, as is required from the observed small spin-orbit interaction in the
non-relativistic quark model.\cite{PGG} In fact, the vector potential is also taken as linear, attractive, 
but without a Coulomb-like contribution, as discussed in Ref.(\cite{Convolve}).

In the LAMP, the confining potentials for each hadron are distributed in an array and are truncated 
on the mid-planes between them. While complex in general, for the case of interest here -- two 
heavy mesons -- the structure is very similar to that of the hydrogen molecule in the Born-Oppenheimer 
approximation, except for the linear vs. inverse distance form of the potential. In this case, the 
large masses of the $c$ and $b$ quarks further enhance the credibility of the approximation -- they 
may be taken in the conventional heavy quark limit~\cite{HQET} as the fixed origins of the confining 
potentials for the light anti-quarks that complete each meson. 

At large separation between the heavy quarks, confinement guarantees the isolation of the light 
quark wave functions from each other. However, as the two mesons approach within a distance 
less than a few times their root-mean-square radii, the truncation of the confining potential allows 
for tunneling of each light anti-quark wave function into the confinement region of the other heavy 
quark rather than the one to which the light anti-quark is initially bound. The concept behind this 
is that a quark can only be confined to nearest center of color attraction, as in string-flip 
models~\cite{stringflip}, for example. This spreading out, or delocalization, of the wave functions 
naturally reduces the localization energy and provides an initial source of binding between the two 
hadrons.

\subsection{Color magnetic and quantum number issues}

In nuclei and other systems, this basic consideration is complicated by additional elements: there 
are color {\bf 6} combinations of quarks and color-magnetic spin interactions of significance on the scale 
of the binding energy. Here again, the concerns raised by these considerations are considerably 
reduced -- the color magnetic interactions between the heavy quarks are reduced by their large 
masses. The light quark  color magnetic interactions with the heavy quarks are also reduced. Only 
the light-quark to light-quark color magnetic interaction remains comparable to that inferred in simple 
quark models of light-quark states. This energy is at most $\approx 50$ MeV as seen~\cite{PDG} in 
individual hadrons (nucleons, $\Delta$'s, light spin-0, and spin-1 mesons) where it depends on the 
color and spin-strong-isospin combinations determined by the constraints of statistics. Furthermore, 
here the presence of both color {\bf 6} and color {\bf 3} combinations, as well as spin-1 and spin-0 
elements, make it clear that strong cancellations of these color magnetic effects to low levels are to 
be expected. Therefore, in this paper we will largely ignore these contributions, since our emphasis 
here is to determine whether the $B$ and $D$ form a four-quark bound state or a more molecular-like 
combination of two identifiable mesons.  We also will neglect the very small electro-magnetic contributions. 
 
Because of these simplifications, in this paper we can also ignore the fact that there are two neutral states 
($B^{-}D^{+}$ and $\bar{B^0}D^0$) that should exist and mix, splitting to form states of definite strong isospin 
(0 and 1) although both have $I_3 = 0$. They also allow us to ignore the detailed spin structures, ranging from 
$J = 0$ to $J = 2$, the last with all of the quark spins aligned.  Also unlike the individual nucleon case, 
the $c$ and $b$ quarks may combine anti-symmetrically to form a color ${\bf {\bar 3}}$ state or symmetrically 
to form a color {\bf  6}.  In the first case, the light anti-quarks must form a color {\bf  3} antisymmetrically, thus 
requiring the spin-isospin combination to be symmetric ($I=0$, $J=0$, or $I=1$, $J=1$) 
and in the latter, the opposite is true -- a color ${\bf {\bar 6}}$ and ($I=1$, $J=0$, or $I=0$, $J=1$). 

Again, these allowed spin-isospin combinations for the light quarks would only produce significant energy 
differences if the color magnetic interaction were larger than the overall binding due to delocalization. The 
color {\bf  6} combination of the heavy quarks would not be expected to produce any attraction, as indeed 
no such components appear in baryons, but the color ${\bf {\bar 3}}$ combination would.  Neither of these 
effects is included here as the channel to color neutralization by decomposition into two color-singlet mesons 
($B$ and $D$) is almost open, so overall color confinement issues should not be significant.
In any event, symmetrization and antisymmetrization between the $c$ and $b$ quarks is moot as they are 
distinguishable. 

We turn now to the detailed calculations of the light (anti)quark wave functions in the double well defined by 
the Born-Oppenheimer-fixed heavy quarks.

\section{The Two-Well Wave Function}

\begin{figure} 
\includegraphics[width=0.6\textwidth, height=0.7\textwidth, angle=0]{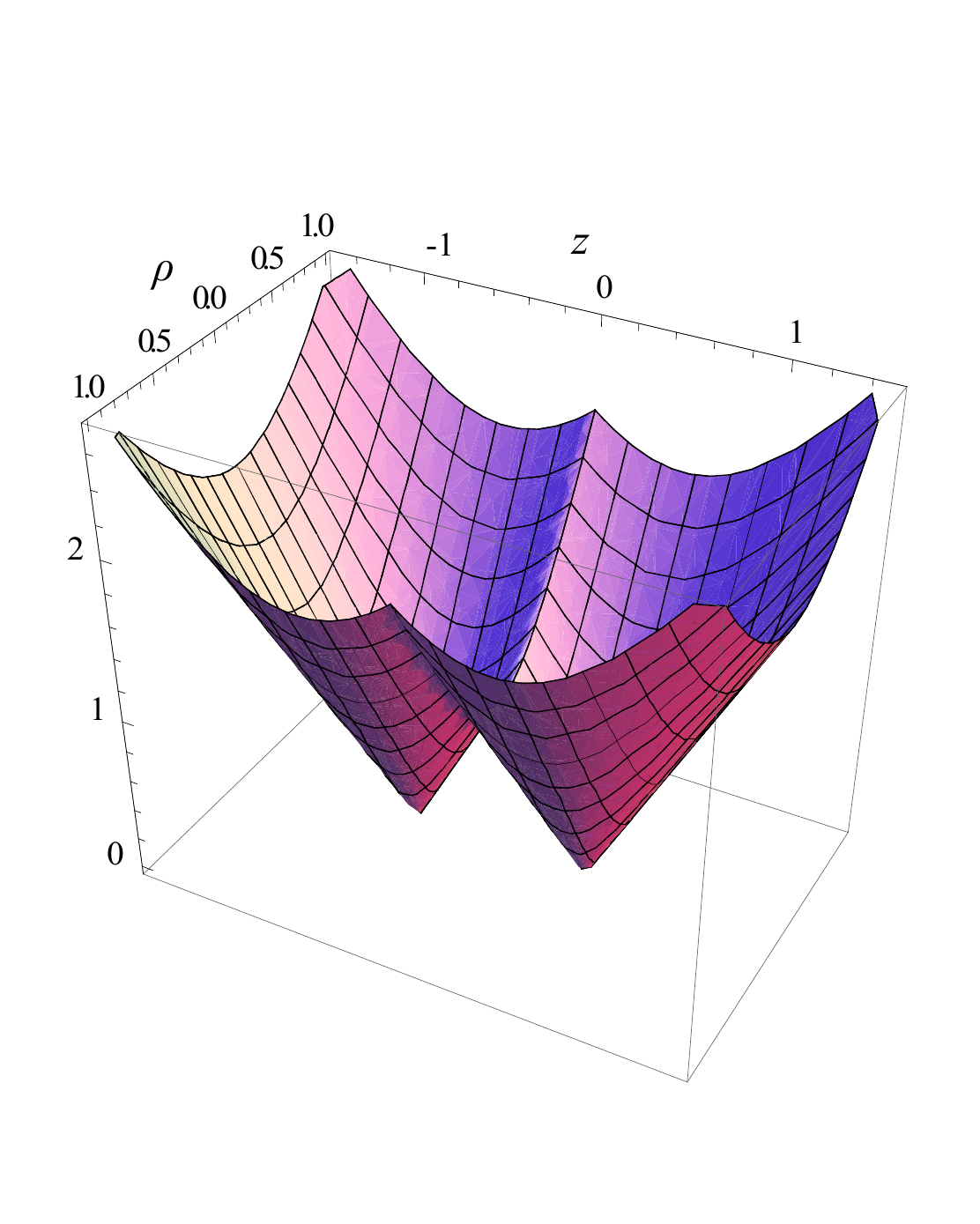}
\caption{Two-well linear potential.  In this and all the following figures, distances,
energies, and wave functions are dimensionless.}
\label{fig:Cones}
\end{figure}
For two wells separated by $2 \delta$ at dimensionless positions
${\bf w}_\pm = \{0,\;0,\;\pm\,\delta\}$ along the $z$-axis (see Fig.\ \ref{fig:Cones}), 
we define the wave function
\Be 
	\Psi_L({\bf r}) = \psi({\bf r}_-) + \epsilon \; \psi({\bf r}_+) , \quad
	\text{ where } \quad {\bf r}_\pm  = {\bf r}  + {\bf w}_\pm  = \{x,y,z\pm\delta\} \ .
	\label{defPsi}
\Ee
This represents, for example, a light $\bar u$-quark (which we assume to be  massless) 
mostly moving and confined in the well at $r_-$ (the ``right'') provided by the heavy $b$-quark.
There may be some ``leakage,'' represented by $\epsilon$, into the ``left'' well
at $r_+$, provided by the heavy $c$-quark.
As mentioned above, we
assume that the $b$ and $c$ quark masses are large enough to justify a 
Born-Oppenheimer approximation of this sort.
There is a similar wave function $\Psi_R$ with $r_-$ and $r_+$ interchanged in 
Eq.\ (\ref{defPsi}) for a light $\bar d$-quark mostly confined to the well at $r_+$ 
with $\epsilon$-leakage into the well at $r_-$.

We will determine variationally what the best values of the parameters $\delta$ and $\epsilon$ 
are that provide a four-quark or molecular-like binding that form a $b \,\bar{u} \;c \,\bar{d}$ system.
The $b$ and $c$ are well separated compared with their Compton sizes.
Since they have little, if any, wave function overlap
and have distinct quantum numbers, anti-symmetrization issues are irrelevant.
For the rest of the paper we will drop the subscripts $L$ and $R$ on $\Psi$, but it should be borne
in mind when we finally compose the $b \,\bar{u} \;c \,\bar{d}$ four-quark state.

In this paper we work as much as possible with dimensionless quantities
(with $\hbar = c = 1$).  
That is, $\delta$,  ${\bf r}$, etc., are all dimensionless distances.
The dimensionless potentials $V(r)$ and $S(r)$ given below in Eq.\ (\ref{onewellVS}) 
are related to dimension-full potentials $\cal V$ and $\cal S$ by a factor of $\kappa^2$,
which has dimensions of GeV/fm.  
For example, $\cal S$ would be defined as
${\cal S}({\sf r}) = \kappa^2 \; {\sf r}$, where ${\sf r} = r/\kappa$ has dimensions in fm.
In GMSS \cite{GMSS}, to cite one reference, $\kappa^2$ was chosen to be 0.9 GeV/fm, 
corresponding to $\kappa = 2.21$ fm$^{-1}$.
In this paper we have used a larger value, $\kappa^2$ = 1.253 GeV/fm, or $\kappa = 2.520$ fm$^{-1}$,
as found in our fitting of charmonia masses.\cite{Convolve}

We take the $\psi$'s in Eq.\ (\ref{defPsi}) to be dimensionless four-component Dirac wave functions 
for light massless $u$- and $d$-quarks.
They are solutions of  
\Be
	H_D \;\psi = [-i\mbox{\boldmath $\alpha$} \cdot {\bf \nabla} + V({\bf r}) + \beta S({\bf r})]\; \psi
	= E \;\psi \ . \label{singleHpsi}
\Ee
Here $V({\bf r})$ is the time component of a Lorentz four-vector and $S({\bf r})$ is a 
Lorentz scalar potential (both to be specified below).
With the Pauli spinor $\chi$ assumed to be quantized along the $z$-direction with 
spin-projection $m_s$, the normalized four-component $s$-wave Dirac wave function 
$\psi({\bf r})$ is
\Be
 \psi_{m_s}({\bf r}) = \frac{1}{\sqrt{4\pi}} \left( \begin{array}{c}
          \psi_a(r)\; \chi_{m_s} \\
	   i  \mbox{\boldmath $\sigma$} \cdot {\bf r} \;\psi_b(r) \; \chi_{m_s}
	   \end{array} \right)  \ . \label{psi}
\Ee

The upper and lower radial wave functions $\psi_a(r)$ and $\psi_b(r)$ can be chosen real.
We have calculated them by solving the coupled radial Dirac equations \cite{EJP} for 
(dimensionless) linear Lorentz vector and scalar potentials of the form
\Be
	V(r) = r - R \quad \text{ and } \quad S(r) = r \ . \label{onewellVS}
\Ee
Here $-R\,$ is a negative displacement pushing the vector potential $V(r)$ down
below the scalar potential $S(r)$, so that confinement trumps Klein-Gordon pair creation.\cite{PGG}

\begin{figure} 
\includegraphics[width=0.6\textwidth, height=0.4\textwidth, angle=0]{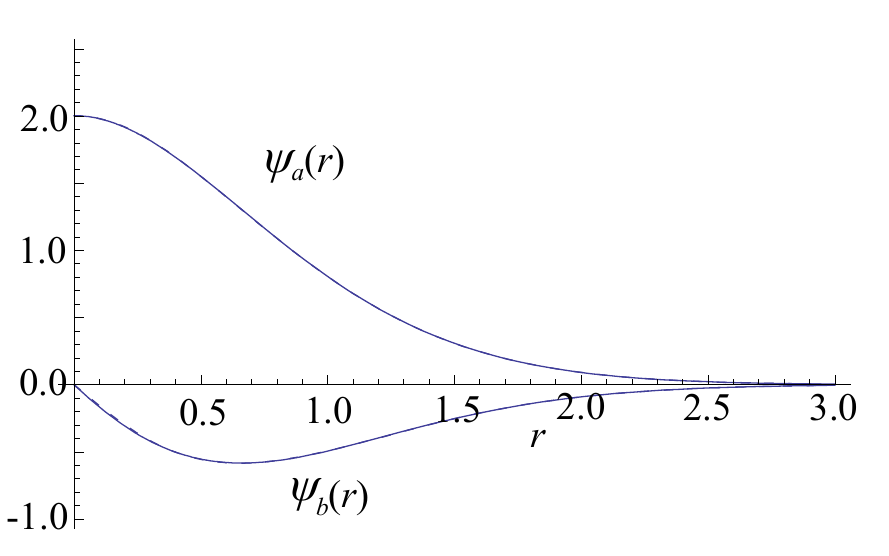}
\caption{Normalized massless quark $1s$ wave functions $\psi_a(r)$ (above the axis) and $r\psi_b(r)$ (below).}
\label{fig:psiab}
\end{figure}
The curves in Fig.\ \ref{fig:psiab} show the calculated (dimensionless) 
$1S$ wave functions $\psi_a(r)$ and $r\psi_b(r)$ 
when the potentials have $R = 1.92$, $\kappa^2 = 1.253$ GeV/fm.
Physical dimensions can be obtained by dividing the dimensionless $r$, $R$, etc., 
by $\kappa = 2.52$ fm$^{-1}$.
The ground state eigenenergy resulting from this calculation is 0.375 GeV.
These potentials provide a reasonable fit to the $c\,\bar{c}$ spectrum.\cite{Convolve}

\section{Expanding $\left<H^{\ 2}_D\right>$}

The idea is that we will want to minimize the expectation value $\left<\,H^2_D\,\right>^{1/2}$ 
with respect to the 
parameters $\epsilon$ and $\delta$ to bound (approximately) the energy for the
four-quark system consisting of $b$, $c$, $\bar{u}$, and $\bar{d}$.
The square $\left<\,H^{\ 2}_D\,\right>$ is required for a variational bound as, 
due to negative energy states, $\left<\,H_D\,\right>$ itself is unbounded below.
The Dirac Hamiltonian $H_D$ is displayed in Eq.\ (\ref{singleHpsi})
but now, for the two-well case (Fig.\ \ref{fig:Cones}), the potentials are
\Be
	V({\bf r}) = \left\{ \begin{array}{ll}
		r_- - R, & \mbox{  if $z > 0$} \\
		r_+ - R, & \mbox{  if $z < 0$} 
						\end{array} 
				\right.			 
\qquad \mbox{ and } \qquad
	S({\bf r}) = \left\{ \begin{array}{ll}
		r_-, & \mbox{  if $z > 0$} \\
		r_+, & \mbox{  if $z < 0$} 
						\end{array} 
				\right.		\ . \label{VStwowell}
\Ee
As already mentioned, $-R\,$ is a negative offset so the vector potential lies below the scalar.

The exact {\it two}-well energy $E$ is in principle found by solving for the eigenvalue of
\Be
	H_D \;\Psi({\bf r}) = E \;\Psi({\bf r}) \ ,
\Ee
with $\Psi$ given in Eq.\ (\ref{defPsi}).
This being difficult, we instead chose to find an approximate value of
the four-quark energy $E$ by the above-mentioned minimization of $\left<H^2_D\right>^{1/2}$.

After some algebra one finds
\Bea
	H^2_D &=& - \nabla^2 + V^2({\bf r}) + S^2({\bf r}) + 2\beta\,V({\bf r})\,S({\bf r}) \nonumber \\
		& &\quad -i\mbox{\boldmath $\alpha$} \cdot \left[ \left({\bf \nabla} V({\bf r})\right) 
			+  \beta\left({\bf \nabla} S({\bf r})\right) \right] 
		-2 i \,V({\bf r}) \;\mbox{\boldmath $\alpha$} \cdot {\bf \nabla} \ . \label{H2D}
\Eea
The lack of a term like $-2 i \,S({\bf r}) \;\mbox{\boldmath $\alpha$} \cdot {\bf \nabla}$
is because of a cancellation (the Dirac operators $\mbox{\boldmath $\alpha$}$ and 
$\beta$ anti-commute).
The first four terms of $H^2_D$ are ``diagonal'' (generically, ${\cal O}_D$) 
in that they connect $\psi_a$ to $\psi_a$ and $\psi_b$ to $\psi_b$, 
while the last two terms are ``off-diagonal'' (${\cal O}_{OD}$) connecting $\psi_a$ to $\psi_b$.

An Appendix describes, in detail, how we calculate the expectation values of the terms 
in Eq.\ (\ref{H2D}).
For brevity, we now present the numerical results of these calculations for $H_D^{\ \ 2}$
and its components.

\begin{figure} 
\includegraphics[width=0.6\textwidth, height=0.4\textwidth, angle=0]{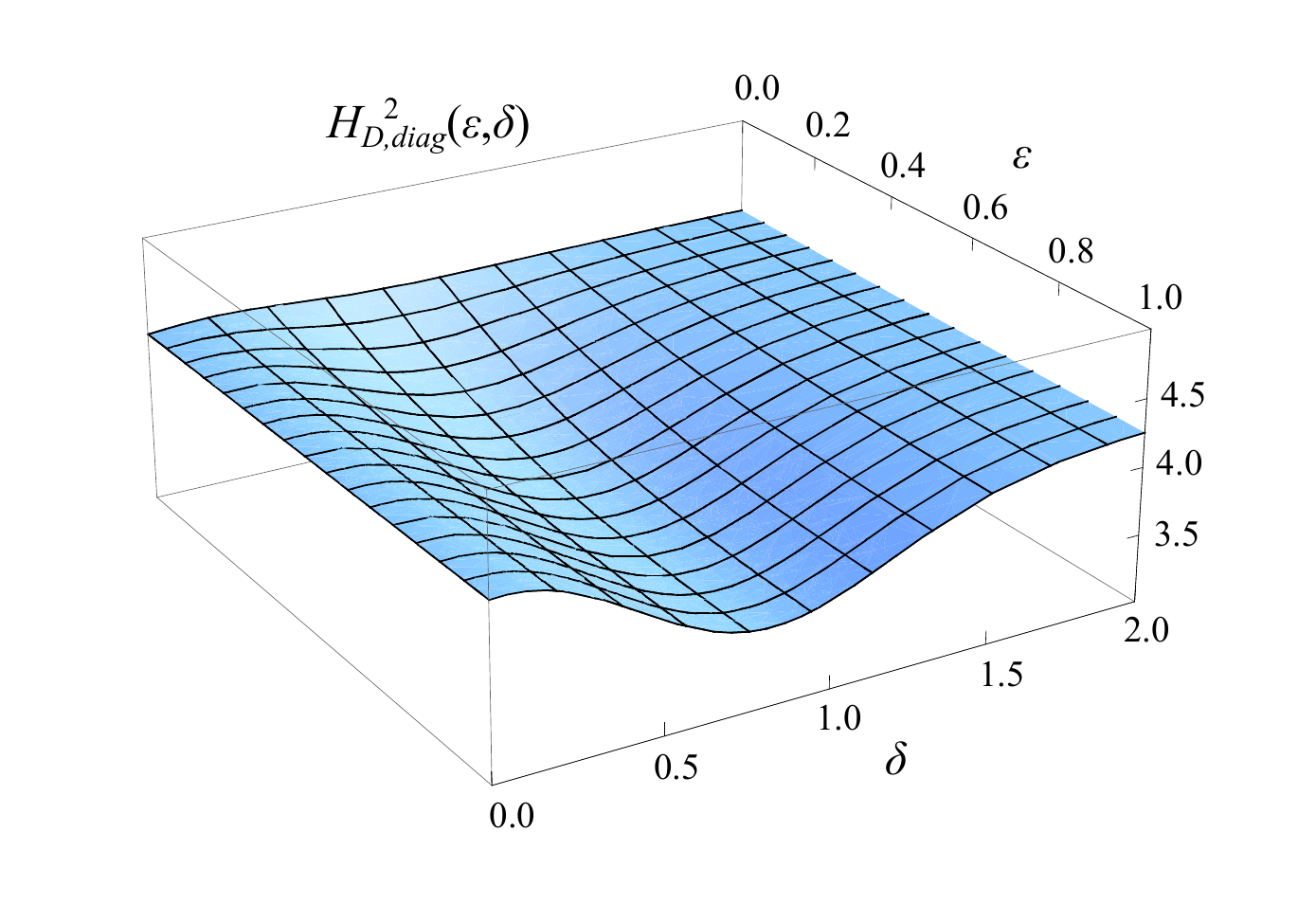}
\caption{Plot of all the diagonal contributions to $<H_D^{\ \ 2}(\epsilon,\delta)>$.}
\label{HD2diagplot}
\end{figure}

\section{Plotting $<H_D^{\ \ 2}>$ to find a minimum energy \label{plotting}}

We combine all the expectation integrals discussed in the Appendix together 
to get an analytic expression for $<H_D^{\ \ 2}>$,
which we can plot to look for a minimum squared energy.

First, we define the (unnormalized) contribution, as a function of $\epsilon$ and $\delta$,
from the diagonal pieces,
\Bea
	<H_{D,\;{\rm diag}}^{\ \ 2}(\epsilon,\delta)> &=&  \sum_{i,j} a_i\, a_j \, 
			\left[\,(1 + \epsilon^2)\,\left( I_{<\nabla^2>}^{(0)} + 
			4\,I_{ij,<r_\pm^2>}^{(0)} - 4\,R\,I_{ij,<r_\pm>}^{(0)} + R^2\,I_{ij,<1>}^{(0)} \right)\right. \nonumber\\
		& & \left. \qquad\qquad\qquad + \,\epsilon \,\left(I_{ij,<\nabla^2>}^{(1)} + 
			4\,I_{ij,<r_\pm^2>}^{(1)} - 4\,R\,I_{ij,<r_\pm>}^{(1)} + R^2\,I_{ij,<1>}^{(1)} \right)\,\right]  \nonumber \\
		& & \!\!\!\!\!\!\!\!\!\!\!\!\!\!\!\!\!\!\!\!\!\!\!\! + \sum_{i,j} b_i\, b_j \, 
			\left[\,(1 + \epsilon^2)\,\left( J_{ij,<\nabla^2>}^{(0)} + R^2\,J_{ij,<1>}^{(0)} \right)
			 + \,\epsilon \,\left( J_{ij,<\nabla^2>}^{(1)} + R^2\,J_{ij,<1>}^{(1)} \right) \,\right]
			 \ , \label{HD2diag}
\Eea
using the expressions for the integrals $I$ and $J$ given in the Appendix.

Figure \ref{HD2diagplot} displays a three-dimensional plot of the normalized
$<H_{D,\;{\rm diag}}^{\ \ 2}(\epsilon,\delta)>/N^2(\epsilon,\delta)$,
where $N^2(\epsilon,\delta)$ is also discussed and displayed in the Appendix.
It shows a relatively shallow minimum at $\epsilon = 1$ and $\delta \approx 0.8 $.
Note the large value, a dimensionless squared-energy of $\approx 4$, which must be largely cancelled
by the off-diagonal contributions to achieve a squared-energy similar to that for the one-well case,
$E^2 = 0.5685$.

The off-diagonal (unnormalized) contributions are
\Bea
	<H_{D,\;{\rm off-diag}}^{\ \ 2}(\epsilon,\delta)> &=&  \sum_{i,j} a_i\, b_j \, 
			\left[\,(1 + \epsilon^2)\,\left( K_{ij,<\nabla VS>}^{(0)} + K_{ij,<V\nabla>}^{(0)} \right)
			\right. \nonumber\\
			& & \left. \qquad\qquad\qquad 
			+\; \epsilon \,\left( K_{ij,<\nabla VS>}^{(1)} + K_{ij,<V\nabla>}^{(1)} \right)\,\right]
			\ , \label{HD2offdiag}
\Eea
with integrals $K$ also from the Appendix.
\begin{figure} 
\includegraphics[width=0.6\textwidth, height=0.4\textwidth, angle=0]{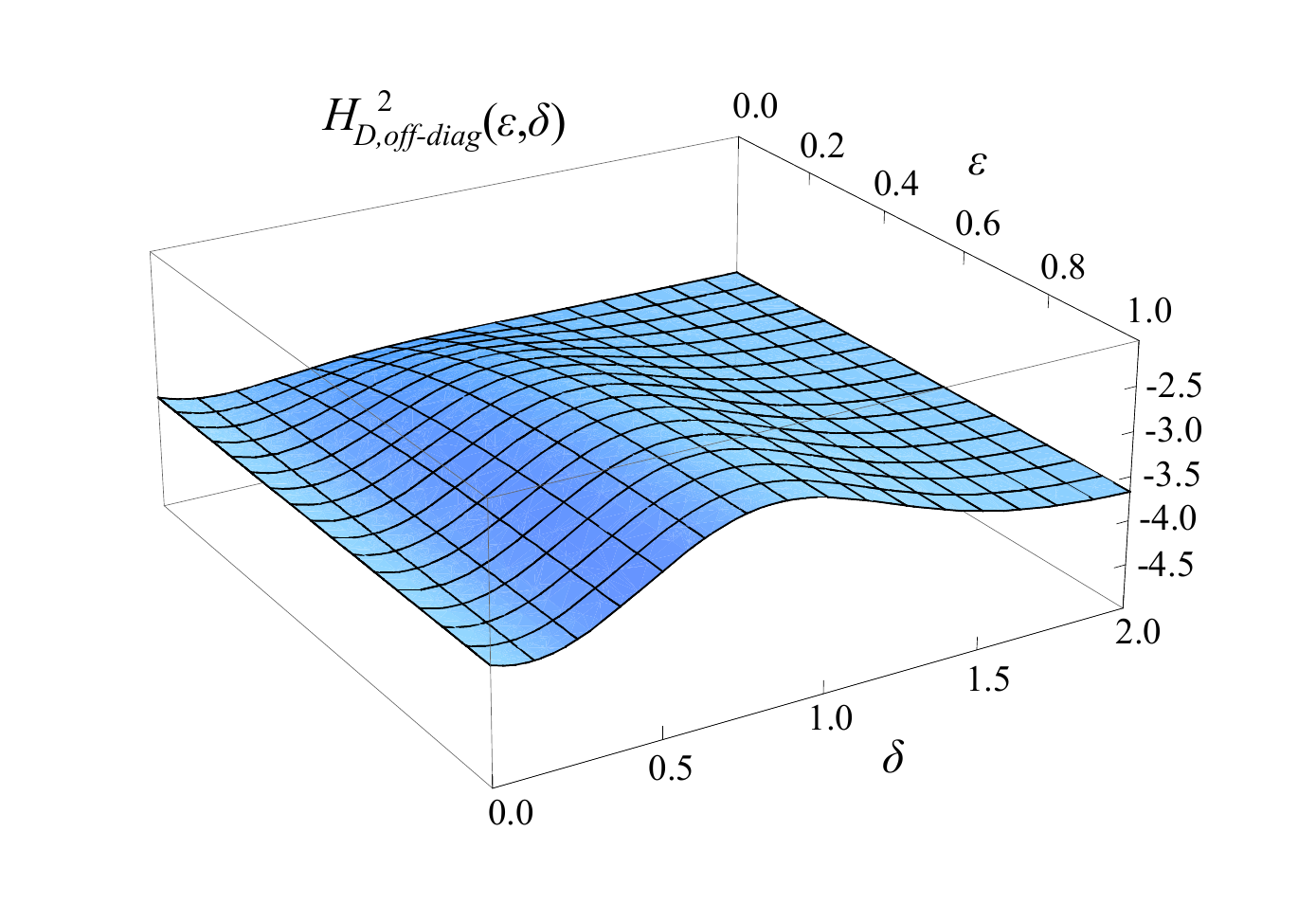}
\caption{Plot of all the off-diagonal contributions to $<H_D^{\ \ 2}(\epsilon,\delta)>$.}
\label{HD2offdiagplot}
\end{figure}
Figure \ref{HD2offdiagplot} gives the three-dimensional plot of 
$<H_{D,\;{\rm off-diag}}^{\ \ 2}(\epsilon,\delta)>/N^2(\epsilon,\delta)$.
In contrast with $<H_{D,\;{\rm diag}}^{\ \ 2}>/N^2$, it has a repulsive hump around $\delta \approx 1$
as well as a shallow valley running from $\epsilon = 0$ to 1 for $\delta \approx 0.2$. 
In the final sum of diagonal and off-diagonal contributions that hump will fill in the minimum
seen in Fig.\ \ref{HD2diagplot}.

Thus we finally combine the two contributions, defining a normalized 
\Be
	<H_{D}^{\ \ 2}(\epsilon,\delta)> = 
		\left[ <H_{D,\;{\rm off-diag}}^{\ \ 2}(\epsilon,\delta)> + 
		<H_{D,\;{\rm off-diag}}^{\ \ 2}(\epsilon,\delta)> \right]/N^2(\epsilon,\delta) \ . \label{HD2}
\Ee
\begin{figure} 
\includegraphics[width=0.6\textwidth, height=0.4\textwidth, angle=0]{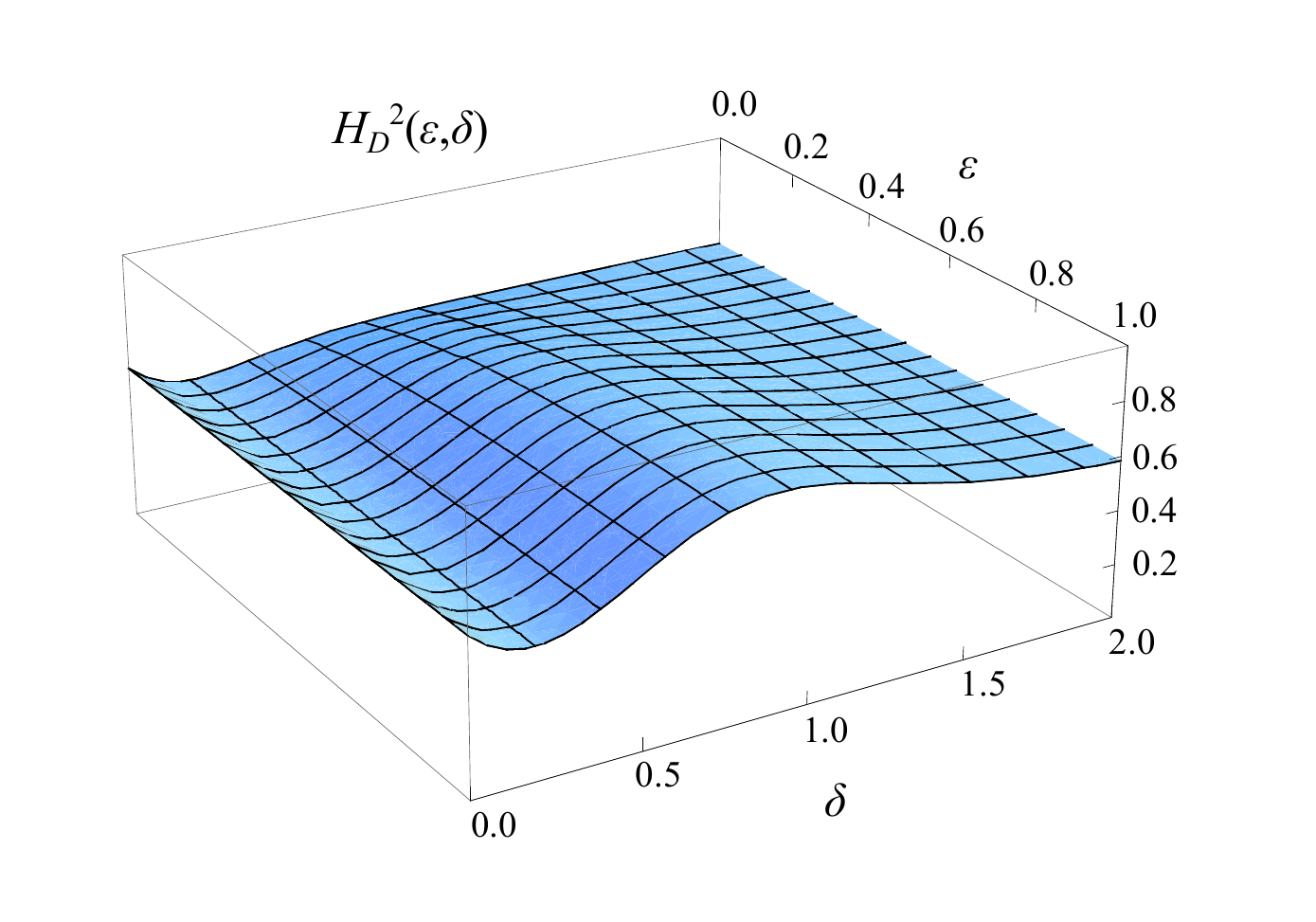}
\caption{Plot of the final $<H_D^{\ \ 2}(\epsilon,\delta)>$.}
\label{HD2plot}
\end{figure}
Figure \ref{HD2plot} plots how $H_{D}^{\ \ 2}$, as a function of $\epsilon$ and $\delta$, develops a
long, flat valley for all values of $\epsilon$ at a separation of $\delta \approx 0.2$ 
(i.e., recalling the value of $\kappa$, a separation of $\approx 0.45$ fm).
Also important is the hump (reminiscent of a fission barrier) around $\delta \approx 0.9$ 
that will help to confine this four-quark system at $\delta \approx 0.2$. 
This hump corresponds to a repulsion between two $Q-\bar{q}$ asymptotic meson states
preventing the light quarks from delocalizing.
There is very little, if any, barrier to coalescence at $\epsilon$ = 0.

\begin{figure} 
\includegraphics[width=0.6\textwidth, height=0.4\textwidth, angle=0]{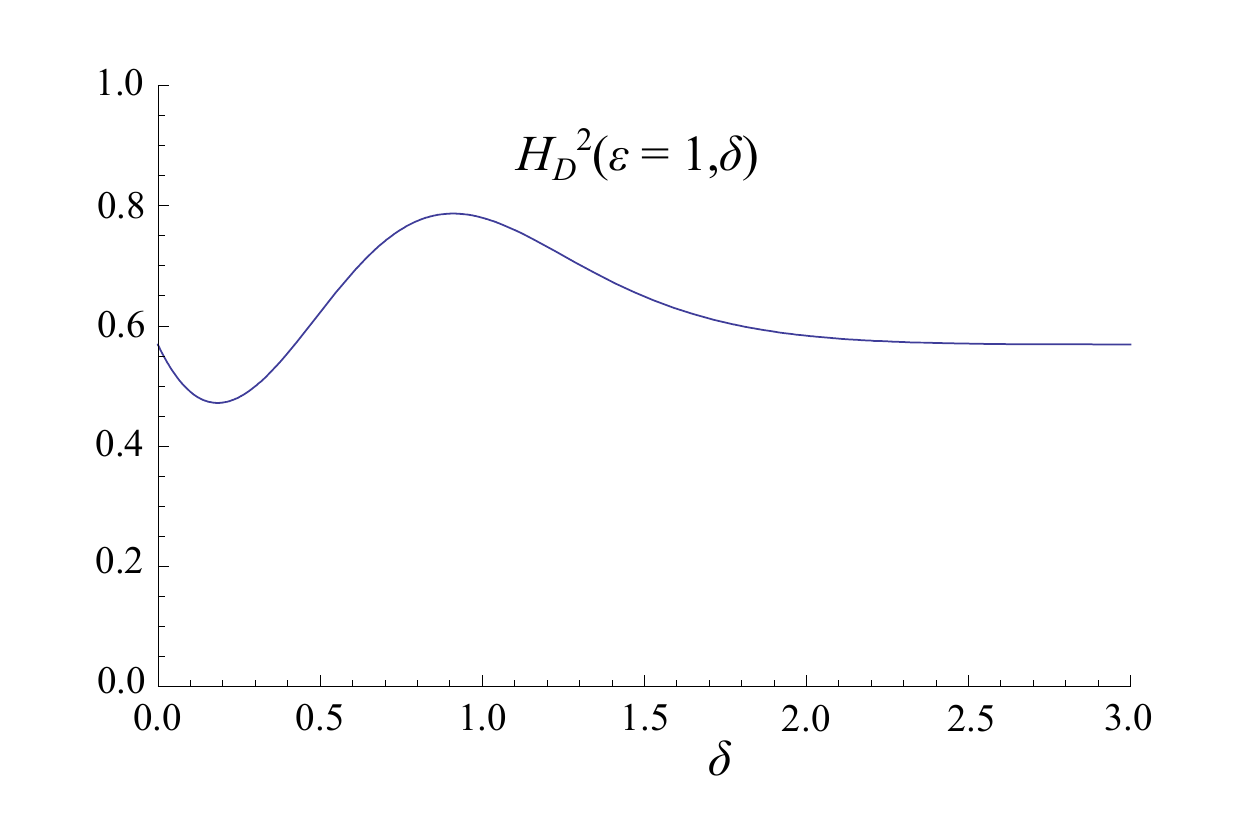}
\caption{$H_D^{\ \ 2}(\epsilon = 1,\delta)$, with a valley at 
$\delta = 0.18$ and a ``fission barrier'' at $\delta \approx 0.9$. }
\label{eps1plot}
\end{figure}
It is easier to see this behavior with a two-dimensional plot, Fig.\ \ref{eps1plot},
showing $H_{D}^{\ \ 2}$ as a function of $\delta$ at
$\epsilon = 1$, where the valley is deepest and the hump is highest.

\begin{figure} 
\includegraphics[width=0.6\textwidth, height=0.4\textwidth, angle=0]{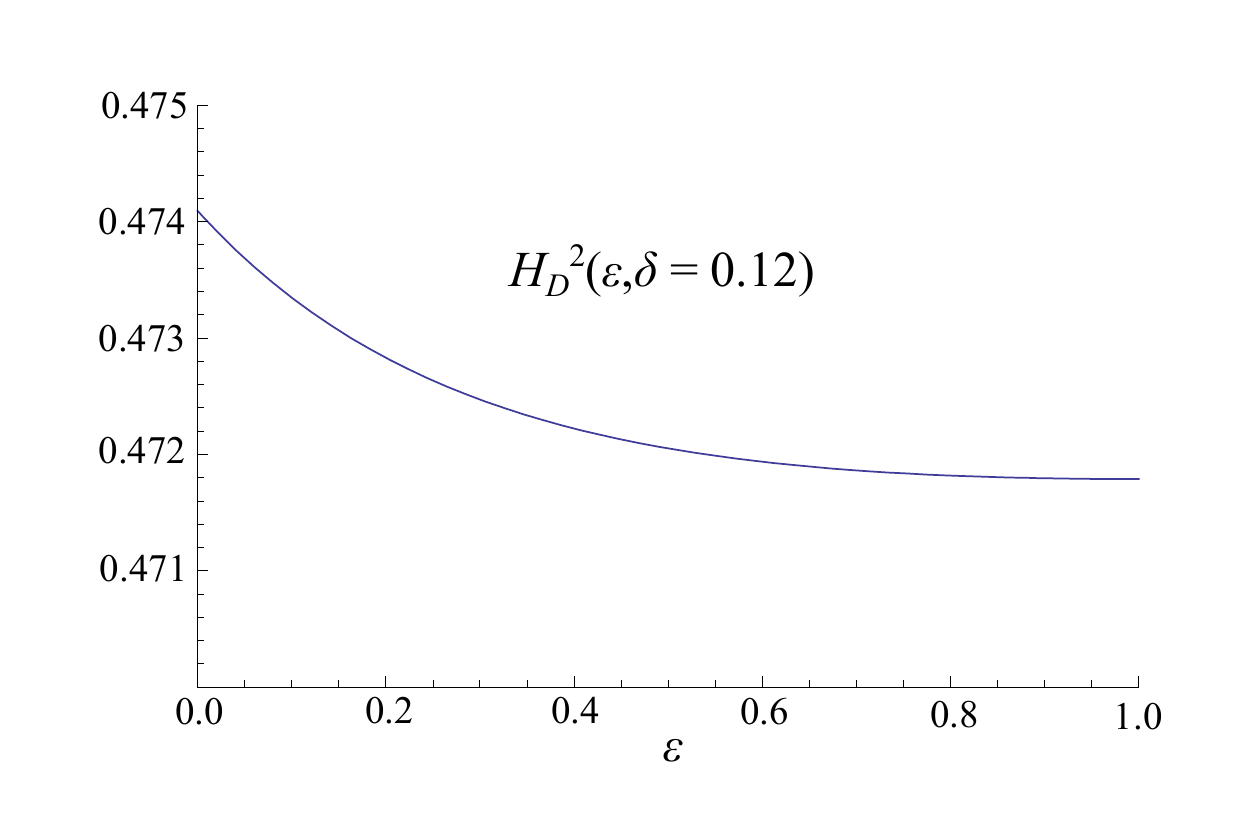}
\caption{Plot of how the nearly flat valley at $\delta = 0.18$ decreases from
$\epsilon = 0$ to $\epsilon = 1$ .}
\label{delta018}
\end{figure}
The dimensionless squared-energy valley-depth at $\epsilon = 1.0$ and $\delta = 0.18$, 
$\Delta H_D^{\ \ 2} = 0.097$, corresponds to a binding energy of 155 MeV for this 
$b\,c\,\bar{u}\,\bar{d}$ four-quark mesonic state.
The valley is surprisingly flat, as shown in Fig.\ \ref{delta018}, 
dropping only 0.0023 squared dimensionless energy units 
from $\epsilon = 0$ to $\epsilon = 1$.
This corresponds to an energy drop of about 24 MeV, a rather small energy difference.
This suggests that Zitterbewegung may play an important role in the nature of this meson.

\section{Discussion}

\begin{figure}[t] 
\includegraphics[width=0.9\textwidth, height=0.9\textwidth, angle=0]{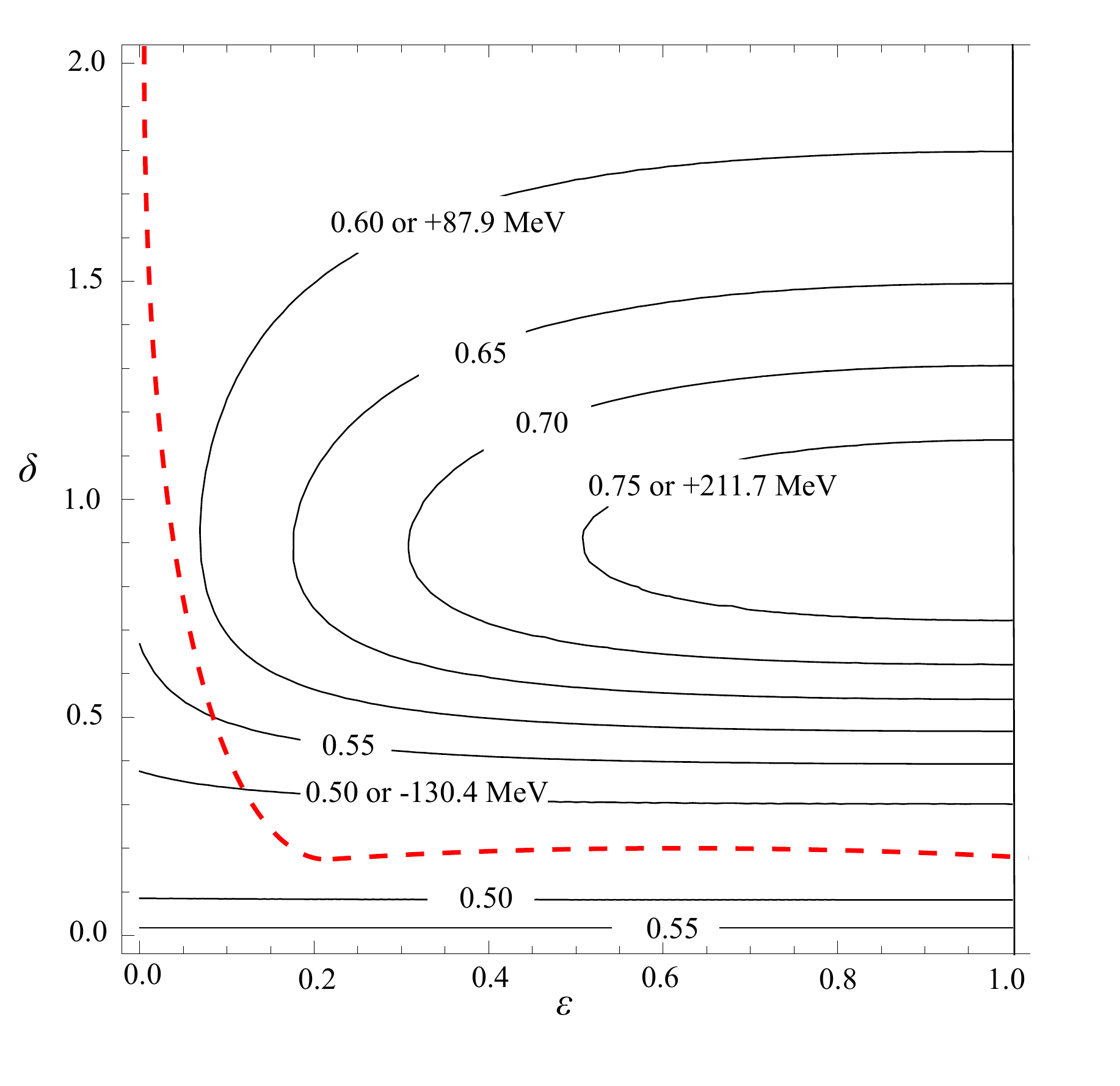}
\caption{Contour plot of $H_D^{\ \ 2}(\epsilon,\delta)$.  The dashed curve illustrates
how two well-separated $Q-\bar{q}$ mesons at $\epsilon = 0$  and large $\delta$ 
come together and slide down the valley
at $\delta \approx 0.2$ to form a four-quark state at $\epsilon = 1$.}
\label{Contour}
\end{figure}

Figure \ref{Contour} is a contour plot of the binding energy of the state in the $\epsilon$-$\delta$ 
plane. It displays two remarkable features: The first is that, at very small $\epsilon$, 
appropriate to the approach towards each other of the two asymptotic ($B$ and 
$D$) mesons, there is no evidence of a repulsive barrier to the fusion of those 
mesons.  The second is that the valley of attraction at small meson separation is 
very flat between small $\epsilon$ ($\sim 0.2$) and $\epsilon = 1$. This indicates 
that there is little energy associated with fluctuations in the $\epsilon$ collective 
variable of the light quarks in the state. There may be a more significant amount 
associated with the $\delta$  collective variable, but this effect is suppressed by 
the large masses associated with the Born-Oppenheimer centers defined by the 
heavy quarks, at least when viewed non-relativistically as seems appropriate
for them, due to their relatively large masses. 
We therefore expect little correction to our estimates of the mass of the four-quark state due to 
collective variable effects. 

The dashed curve in Figure\ \ref{Contour} illustrates how two well-separated $B$ and $D$ mesons at 
$\epsilon = 0$ and large $\delta$ would come together to $\delta \approx 0.2$ and 
$\epsilon \approx 0.2$, corresponding to a heavy quark separation of about $0.45$ fm. 
As we have emphasized above, this small separation makes it clear that long-range 
pion-exchange effects do not contribute significantly. 
From $\epsilon \approx  0.2$, the four-quark 
state then slides gently down the nearly flat valley to $\epsilon = 1$ where it is most bound.
Such a state is prevented from falling apart because of the ``fission barrier'' 
around $\delta \approx 0.9$.

We have ignored the possible color magnetic contributions from the interaction of the two 
light antiquarks, but this must be less than 50 MeV and we expect it to be even less than 
half this value. These corrections, which we will deal with in a future publication,
are not large compared to the extracted variational upper bound on the binding energy 
of order 150 MeV found in our calculations. 
Thus, by comparing our binding energy with the threshold for $B$ and $D$ mesons, 
we predict a set of states in the region of 7 GeV/$c^2$.

Finally, we comment on the surprisingly small difference in binding energy between 
the ``molecular'' form of the bound state, ($\epsilon \approx 0.2$, as in nuclei \cite{GMSS}))
and the four-quark limit ($\epsilon = 1$).
If this feature is widespread in such heavy quark systems, it could go far towards
explaining why it has been so difficult to identify unambiguous four-quark states. 

In any event, as our interest here is in nuclear physics, we note that the small separation 
compared to root-mean-square size of the meson states argues against the identification 
of the system as that of two slightly off-shell free space mesons, at least, at $\epsilon \sim 1$. 
However, the small difference in energy between that  region and $\epsilon \sim 0.2$ suggests 
to the contrary, that since the binding energy is not large, at least some of the time, the system 
would appear to be one described as two slightly off-shell free space mesons, with substantial 
fluctuations between the two pictures. Difficult as it was historically, we conclude that nuclear 
physics would have been even more difficult to understand if it had similar properties. 

\section{Acknowledgments}

This work was carried out in part under the auspices of the National Nuclear Security
Administration of the U.S. Department of Energy at Los Alamos National Laboratory
under Contract No. DE-AC52-06NA25396.

\pagebreak
\appendix
\section{Calculational Details}

\subsection{Approximating $\psi_a$ and $\psi_b$ as a sum of Gaussians}

For the calculations presented below, the $\psi_{a,b}$ have both been fitted to sums of Gaussians, 
\Be 
	\psi_a(r) = \sum_i a_i\; e^{-\mu_i r^2/2}, \quad \psi_b(r) = \sum_i b_i\; e^{-\mu_i r^2/2} \ ,
	\label{gaussians}
\Ee
where the $a_i$, $b_i$, and $\mu_i$ are dimensionless numbers.  
We found it necessary to go to six terms, so that evaluating the upper and lower components 
of the left-hand-side of the Dirac equation [Eq.\ (2) in the main text] gives reasonable agreement with the right-hand-side.
The fitted parameters are
\Bea
	\mu_i &=& \ {\; 1.0,1.3,1.6,2.0,4.0,8.0} \;\} \nonumber \label{abmus} \\
	a_i &=& \{\; 0.492649, -0.687482, 1.84609, -0.00246039, 0.258295, 0.0956581 \;\} \\
	b_i &=& \{\; -0.0571296, 1.03367, -1.18398, 1.33989, 0.162575, 0.299479 \;\} 
		\nonumber \ .
\Eea
The fitted $\psi_a(r)$ and $r\psi_b(r)$ are shown as the dashed curves in Fig.\ \ref{fig:psiab}, 
largely overlying the solid curves from the solution of the Dirac equation.
To check the quality of the fits we have evaluated the single-quark expectation value
of the Hamiltonian, $<\,H_D\,> = 0.7545$, which is slightly larger than the 
(dimensionless) energy eigenvalue $E = 0.7540$ (which, for a variational trial function, 
is as it should be).
As a second check on our Gaussian fits of $\psi_a$ and $\psi_b$, Eqs.\ (\ref{gaussians}) 
and (\ref{onewellVS}), we also evaluated the single-well expectation $<\,H_D^2\,>$ to be 0.5691, 
again slightly larger than $E^2 = 0.5685$, as it should be.

\subsection{General Remarks on calculating the expectations \label{general}}

The reason for approximating our numerical radial wave functions $\psi_a$ and $\psi_b$
as sums of Gaussians is that it allows us to calculate the expectation values of each
of the terms of $H_D^2$ analytically.  
Given an analytic expression for $H_D^2$ allows us to plot it quickly and precisely as 
a function of the variational parameters $\delta$ and $\epsilon$.
To do these integrations, we have relied heavily on programs such as Mathematica and Maple.
As will be seen, the final results can sometimes be messy and often involve error  
functions\footnote{See, e.g., M.\ Abramovitz and I.\ A.\ Stegun, {\it Handbook of 
Mathematical Functions}, (Dover, New York, 1965), Chap.\ 7} because of the Gaussians being integrated. 

For the diagonal operators of $H^2_D$ we will calculate the upper and lower contributions separately,
\Be
	\left< \Psi | {\cal O}_D | {\Psi} \right> = \left< \Psi | {\cal O}_D | \Psi \right>_A + 
		\left< \Psi | {\cal O}_D | \Psi \right>_B \ .
\Ee
The $B$-expectations are more complicated than those for $A$ because of the factors
of $-i\mbox{\boldmath $\sigma$} \cdot {\bf {r_\pm}}$ multiplying the radial $\psi_b$'s.
However, for some diagonal operators, as will be seen below, the $B$-expectations are not always needed.
In any case, from Eq.\ (\ref{gaussians}) we expand these diagonal operator expectations as 
\Be
	 \left< \Psi | {\cal O}_D | \Psi \right>_A = \sum_{i,j} a_i\, a_j \; I_{ij} \ , \qquad
	 \left< \Psi | {\cal O}_D | \Psi \right>_B = \sum_{i,j} b_i\, b_j \; J_{ij} \ , 
	 \label{ODAandB}
\Ee
where the $I_{ij}$ and $J_{ij}$ are integrals over Gaussians.

First, we separate out the quadratic dependence on $\epsilon$ as
\Be
	I_{ij} = I_{ij}^{(0)} + \epsilon \; I_{ij}^{(1)} + \epsilon^2 \; I_{ij}^{(2)}
		= (1 + \epsilon^2) \; I_{ij}^{(0)} + \epsilon \;  I_{ij}^{(1)} \ , \label{epsilonexpansion}
\Ee
and likewise for the lower-component $B$-integrals $J_{ij}$.
The second equality here comes about because parity symmetry ensures that 
the $I_{ij}^{(2)} = I_{ij}^{(0)}$, etc.
We will refer to the $I_{ij}^{(0)}$ as ``direct terms,'' in that they connect Gaussians
with $\mu_j \,r_-^2/2$ to those with $\mu_i \,r_-^2/2$ (and similarly for $I_{ij}^{(2)}$ with $r_+$).
Recalling the $1/4\pi$ from the normalization of the $\psi$'s, we ensure the symmetry under the
interchange of indices $i$ and $j$ by writing
\Be
	I_{ij}^{(0)} = \frac{1}{8\pi}\; \int d^3 r \;  
		\left\{\;e^{-\mu_i \,r^2_-/2}\;{\cal O}_D\;e^{-\mu_j \,r^2_-/2} 
		+ e^{-\mu_j \,r^2_-/2}\;{\cal O}_D\;e^{-\mu_i \,r^2_-/2} \;\right\} \ . \label{genericI0}
\Ee
The direct integrals $J_{ij}^{(0)}$ have a similar form but with
$\mbox{\boldmath $\sigma$} \cdot {\bf r}_- \;{\cal O}_D \; \mbox{\boldmath $\sigma$} \cdot {\bf r}_-$
in place of the ${\cal O}_D$.

The ``cross terms'' $I_{ij}^{(1)}$ are more complicated integrals than the $I_{ij}^{(0)}$,
and likewise for $J_{ij}^{(1)}$.
They connect Gaussians with $\mu_j \,r_-^2/2$ to $\mu_j \,r_+^2/2$ and vice versa.
Thus, on symmetrizing in $i$ and $j$,
\Bea
	I_{ij}^{(1)} &=& \frac{1}{8\pi}\; \int d^3 r \; \left\{ \left[ 
		e^{-\mu_i \,r^2_-/2}\;{\cal O}_D \;e^{-\mu_j \,r^2_+/2} +  
		e^{-\mu_i \,r^2_+/2}\;{\cal O}_D \;e^{-\mu_j \,r^2_-/2} +  \right] \right. \nonumber \\ 
		& & \left. \qquad\qquad + \;\left[ 
		e^{-\mu_j \,r^2_-/2}\;{\cal O}_D \;e^{-\mu_i \,r^2_+/2} + 
		e^{-\mu_j \,r^2_+/2}\;{\cal O}_D \;e^{-\mu_i \,r^2_-/2} \right]	\right\} \label{genericI1}
\Eea
The $J_{ij}^{(1)}$ have a similar form but with ${\cal O}_D$ replaced by
$\mbox{\boldmath $\sigma$} \cdot {\bf r}_- \;{\cal O}_D \; \mbox{\boldmath $\sigma$} \cdot {\bf r}_+ \ $
or 
$\ \mbox{\boldmath $\sigma$} \cdot {\bf r}_+ \;{\cal O}_D \; \mbox{\boldmath $\sigma$} \cdot {\bf r}_-$,
as appropriate.


Each of the off-diagonal operators in Eq.\ (9) of the main text has the general form
\Be
	{\cal O}_{OD} = -i \mbox{\boldmath $\alpha$} \cdot {\bf X} = 
			\left[ \begin{array}{cc}
		 0 & -i\;\mbox{\boldmath $\sigma$} \cdot {\bf X}_{12} \\
		-i\;\mbox{\boldmath $\sigma$} \cdot {\bf X}_{21} & \quad 0 
						\end{array} 
		\ \right] \ \label{Xij}
\Ee
where the ${\bf X}_{12}$ and ${\bf X}_{21}$ are vector-operators that may not be equal 
because of the possible presence of the diagonal $\beta$ matrix in ${\cal O}_{OD}$.

The direct terms of the off-diagonal expectation $<{\cal O}_{OD}>$ involve several terms because 
the upper component of $\Psi^\dagger({\bf r}_-)$ connects through 
$-i\;\mbox{\boldmath $\sigma$} \cdot {\bf X}_{12}$ to the lower component of $\Psi({\bf r}_-)$ 
at the same time that the lower component of $\Psi^\dagger({\bf r}_-)$ connects through 
$-i\;\mbox{\boldmath $\sigma$} \cdot {\bf X}_{21}$ to the upper component of $\Psi({\bf r}_-)$.
We therefore have to keep the sums over the $a$'s and $b$'s in Eq.\ (\ref{gaussians}) as
parts of the integrand.
Again symmetrizing in $i$ and $j$,
\Bea
	<{\cal O}_{OD}^{(0)}> &=& \frac{1}{8\pi}\; \sum_{i,j}  \int d^3 r \; \left\{ 
			e^{-\mu_i \,r^2_-/2}\; \left[ -a_i b_j 
			(\mbox{\boldmath $\sigma$} \cdot {\bf X}_{12})
			(\mbox{\boldmath $\sigma$} \cdot {\bf r}_-) \right. \right. \nonumber \\
		& & \qquad\qquad\qquad\qquad\qquad \left. +\; a_j b_i 
			(\mbox{\boldmath $\sigma$} \cdot {\bf r}_-) 
			(\mbox{\boldmath $\sigma$} \cdot {\bf X}_{21}) 
			\right] \; e^{-\mu_j \,r^2_-/2}  \nonumber \\
		& & \qquad\qquad\quad \left.  +\;
			e^{-\mu_j \,r^2_-/2}\; \left[ -a_j b_i 
			(\mbox{\boldmath $\sigma$} \cdot {\bf X}_{12}) 
			(\mbox{\boldmath $\sigma$} \cdot {\bf r}_-) \right. \right. \nonumber \\
		& & \qquad\qquad\qquad\qquad\qquad \left. \left. +\; a_i b_j 
			(\mbox{\boldmath $\sigma$} \cdot {\bf r}_-)  
			(\mbox{\boldmath $\sigma$} \cdot {\bf X}_{21}) 
			\right] \; e^{-\mu_i \,r^2_-/2}    
			\right\} \ . \label{formOOD0}
\Eea

The cross terms of $<{\cal O}_{OD}>$ have even more terms because the $\Psi^\dagger({\bf r}_+)$ 
connects to $\Psi({\bf r}_-)$ at the same time that $\Psi^\dagger({\bf r}_-)$ connects 
to $\Psi({\bf r}_+)$.
It becomes
\Bea
	<{\cal O}_{OD}^{(1)}> &=& \frac{1}{8\pi}\; \sum_{i,j}  \int d^3 r \; \left\{ 
			e^{-\mu_i \,r^2_+/2}\; \left[ -a_i b_j 
			(\mbox{\boldmath $\sigma$} \cdot {\bf X}_{12})
			(\mbox{\boldmath $\sigma$} \cdot {\bf r}_-) \right. \right. \nonumber \\
		& & \qquad\qquad\qquad\qquad\qquad \left. +\; a_j b_i 
			(\mbox{\boldmath $\sigma$} \cdot {\bf r}_+) 
			(\mbox{\boldmath $\sigma$} \cdot {\bf X}_{21}) 
			\right] \; e^{-\mu_j \,r^2_-/2}  \nonumber \\
		& & \qquad\qquad\quad \left.  +\;
			e^{-\mu_i \,r^2_-/2}\; \left[ -a_i b_j 
			(\mbox{\boldmath $\sigma$} \cdot {\bf X}_{12})
			(\mbox{\boldmath $\sigma$} \cdot {\bf r}_+) \right. \right. \nonumber \\
		& & \qquad\qquad\qquad\qquad\qquad \left. +\; a_j b_i 
			(\mbox{\boldmath $\sigma$} \cdot {\bf r}_-) 
			(\mbox{\boldmath $\sigma$} \cdot {\bf X}_{21}) 
			\right] \; e^{-\mu_j \,r^2_+/2} \nonumber \\
		& & \qquad\qquad\quad \left.  +\;
			e^{-\mu_j \,r^2_+/2}\; \left[ -a_j b_i 
			(\mbox{\boldmath $\sigma$} \cdot {\bf X}_{12}) 
			(\mbox{\boldmath $\sigma$} \cdot {\bf r}_-) \right. \right. \nonumber \\
		& & \qquad\qquad\qquad\qquad\qquad \left.  +\; a_i b_j 
			(\mbox{\boldmath $\sigma$} \cdot {\bf r}_+)  
			(\mbox{\boldmath $\sigma$} \cdot {\bf X}_{21}) 
			\right] \; e^{-\mu_i \,r^2_-/2}  \nonumber \\
		& & \qquad\qquad\quad \left.  +\;
			e^{-\mu_j \,r^2_-/2}\; \left[ -a_j b_i 
			(\mbox{\boldmath $\sigma$} \cdot {\bf X}_{12})
			(\mbox{\boldmath $\sigma$} \cdot {\bf r}_+) \right. \right. \nonumber \\
		& & \qquad\qquad\qquad\qquad\qquad \left. \left. +\; a_i b_j 
			(\mbox{\boldmath $\sigma$} \cdot {\bf r}_-) 
			(\mbox{\boldmath $\sigma$} \cdot {\bf X}_{21}) 
			\right] \; e^{-\mu_j \,r^2_+/2}   
			\right\} \ . \label{formOOD1}
\Eea

The integrations for the $I$'s, $J$'s, and in Eqs.\ (\ref{formOOD0}) and (\ref{formOOD1}) 
can best be done using (dimensionless) 
cylindrical coordinates, $\rho = \left({x^2 + y^2}\right)^{1/2}$, $\theta$, and $z$.
The $\theta$ integrations are trivial, providing a factor of $2\pi$, which will cancel with 
the $1/4\pi$ coming from the normalizations of the $\psi$'s in Eq.\ (\ref{psi})
to give an overall factor of $1/2$ before each double integral over $\rho$ and $z$.
It usually is easier to do the $\rho$-integration (from 0 to $+\infty$) first.
Because $V({\bf r})$ and $S({\bf r})$ depend on $r_-$ when $z > 0$ 
and on $r_+$ when $z < 0$, we need to do the $z$-integration separately for those regions,
i.e., for $z$ from $-\infty$ to 0 and then for $z$ from 0 to $+\infty$.  
The separate results are then added and simplified to give the final integral.

We will distinguish the results for the expectations of the different operators 
in Eq.\ (\ref{H2D}) by an appropriate subscript.
For example, for $O_D = \nabla^2$, we will write $I_{ij}^{(0,1)}$ as 
$I_{ij,\;<\nabla^2>}^{(0,1)}$, and similarly for the $J_{ij}$ integrals.

\subsection{Normalizing $\Psi$ \label{normPsi}}

While the Dirac $\psi$'s are themselves properly normalized, the two-well $\Psi$ is not.
For this we need to calculate the expectation values of ${\cal O}_D = 1$ to find
\Be
	N^2(\delta,\epsilon) = \int d^3 r \; \Psi^\dagger \Psi = \left< \Psi | 1 | \Psi \right> = 
		\left< \Psi | 1 | \Psi \right>_A + \left< \Psi | 1 | \Psi \right>_B \ .
\Ee  
We make the expansion in $\epsilon$ as in Eq.\ (\ref{epsilonexpansion}) above.
The direct-term integrals for the expectation $\left< \,1\, \right>$ are,
noting that for the $J_{ij,\;<1>}^{(0)}$ we also have a factor of
$(\mbox{\boldmath $\sigma$} \cdot {\bf r}_-) (\mbox{\boldmath $\sigma$} \cdot {\bf r}_-) = r_-^2$
in the integrand,

\Bea
	I_{ij,\;<1>}^{(0)} 
		&=& \left[\frac{\pi}{2(\mu_i + \mu_j)^3}\right]^{1/2} \\
	J_{ij,\;<1>}^{(0)} 
		&=& 3 \left[\frac{\pi}{2(\mu_i + \mu_j)^5}\right]^{1/2} \ ,
\Eea
both independent of $\delta$.

The cross-term integrals do depend on $\delta$. 
For the $J_{ij,\;<1>}^{(1)}$ we need the factor
\Be
	(\mbox{\boldmath $\sigma$} \cdot {\bf r}_+) (\mbox{\boldmath $\sigma$} \cdot {\bf r}_-) 
	= {\bf r}_+ \cdot {\bf r}_- = \rho^2 + z^2 - \delta^2 \  
\Ee
in the integrand.
Proceeding as in Sec.\ \ref{general}, we find
\Bea
	I_{ij,\;<1>}^{(1)} 
			&=& \left[\frac{2\pi}{(\mu_i + \mu_j)^3}\right]^{1/2}\,  
						e^{-2 \mu_i\mu_j\;\delta^2/(\mu_i + \mu_j)} \ , \label{I1one} \\
	J_{ij,\;<1>}^{(1)} 
			&=& \left[\; 3(\mu_i + \mu_j) - 4 \;\mu_i \mu_j\;\delta^2 \;\right]\;
				\left[\frac{2\pi}{(\mu_i + \mu_j)^7}\right]^{1/2}\, 
				e^{-2 \mu_i\mu_j\;\delta^2/(\mu_i + \mu_j)} \ .
\Eea
Note that, when $\delta = 0$, $I_{ij,\;<1>}^{(1)} = 2 \; I_{ij,\;<1>}^{(0)}$, 
and $J_{ij,\;<1>}^{(1)} = 2 \; J_{ij,\;<1>}^{(0)}$.
This is a common feature for all the expectations here and below.
This is necessary so that, for example, when $\delta = 0$ and $\epsilon = 1$, 
one recovers a result that is four times that when $\delta = 0$ and $\epsilon = 0$.

\begin{figure} 
\includegraphics[width=0.8\textwidth, height=0.35\textwidth, angle=0]{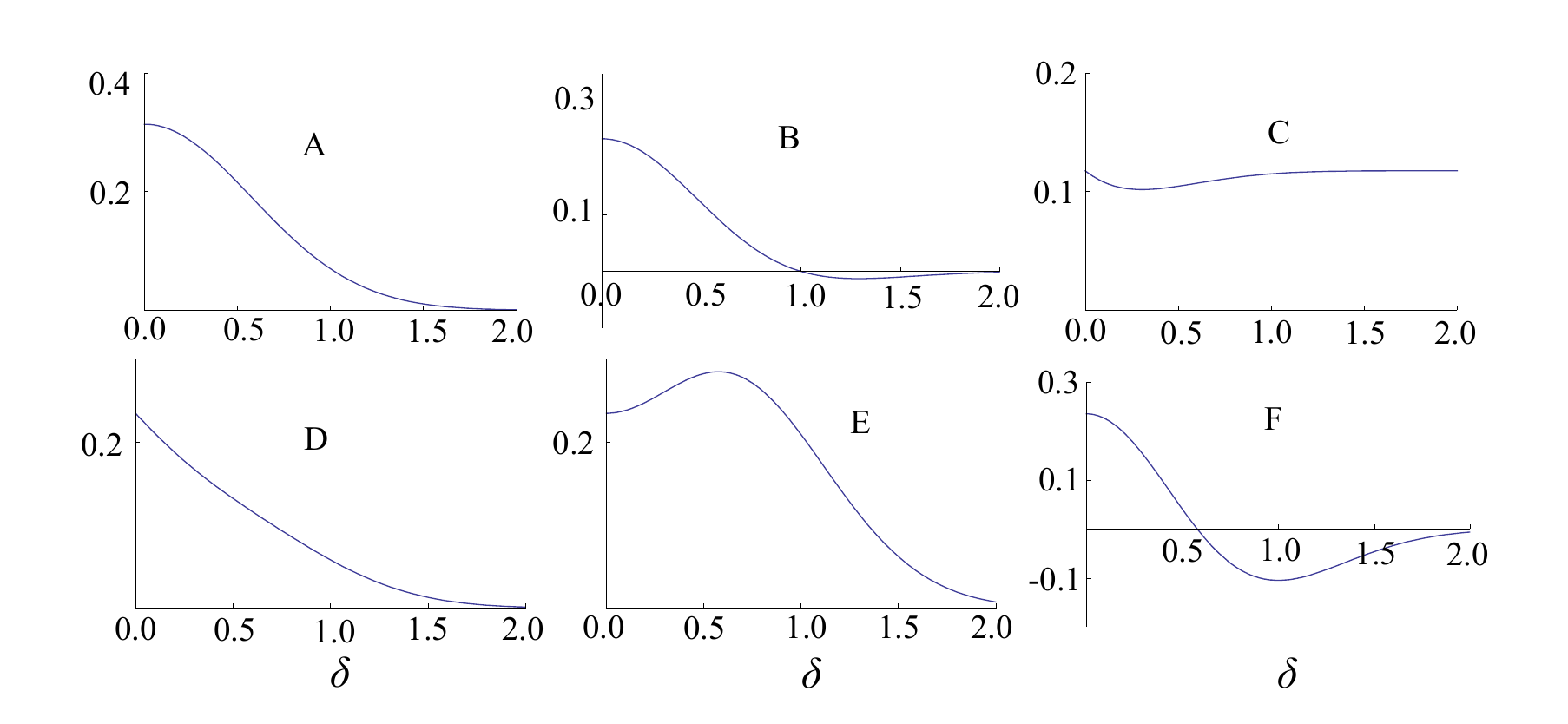}
\caption{Typical plots of $I$'s, $J$'s, and $K$'s as functions of $\delta$.
The $y$-axes are in arbitrary units.}
\label{sixplots}
\end{figure}
We see from Eq.\ (\ref{I1one}) that $I_{ij,\;<1>}^{(1)}$,as a 
function of $\delta$, is a decaying Gaussian (as in Fig.\ \ref{sixplots}, plot A).
On the other hand, $J_{ij,\;<1>}^{(1)}$ falls off from its peak at $\delta = 0$, goes through zero, and 
has a mild minimum before decaying to zero at large $\delta^2$ (as in Fig.\ \ref{sixplots},
plot B).
 

Combining all terms,
\Bea
	N^2(\epsilon,\delta) &=& \sum_{i,j} a_i\, a_j \, 
		\left[\,(1 + \epsilon^2)\,I_{<1>}^{(0)} + \epsilon \,I_{<1>}^{(1)}\,\right] + 
		\sum_{i,j} b_i\, b_j \, \left[\,(1 + \epsilon^2)\, J_{<1>}^{(0)} + 
		\epsilon \,J_{<1>}^{(1)}\,\right]
\Eea
and the normalized $\Psi$ is obtained by multiplying Eq.\ (\ref{defPsi}) by 
${1/N(\epsilon,\delta)}$.

\begin{figure} 
\includegraphics[width=0.6\textwidth, height=0.5\textwidth, angle=0]{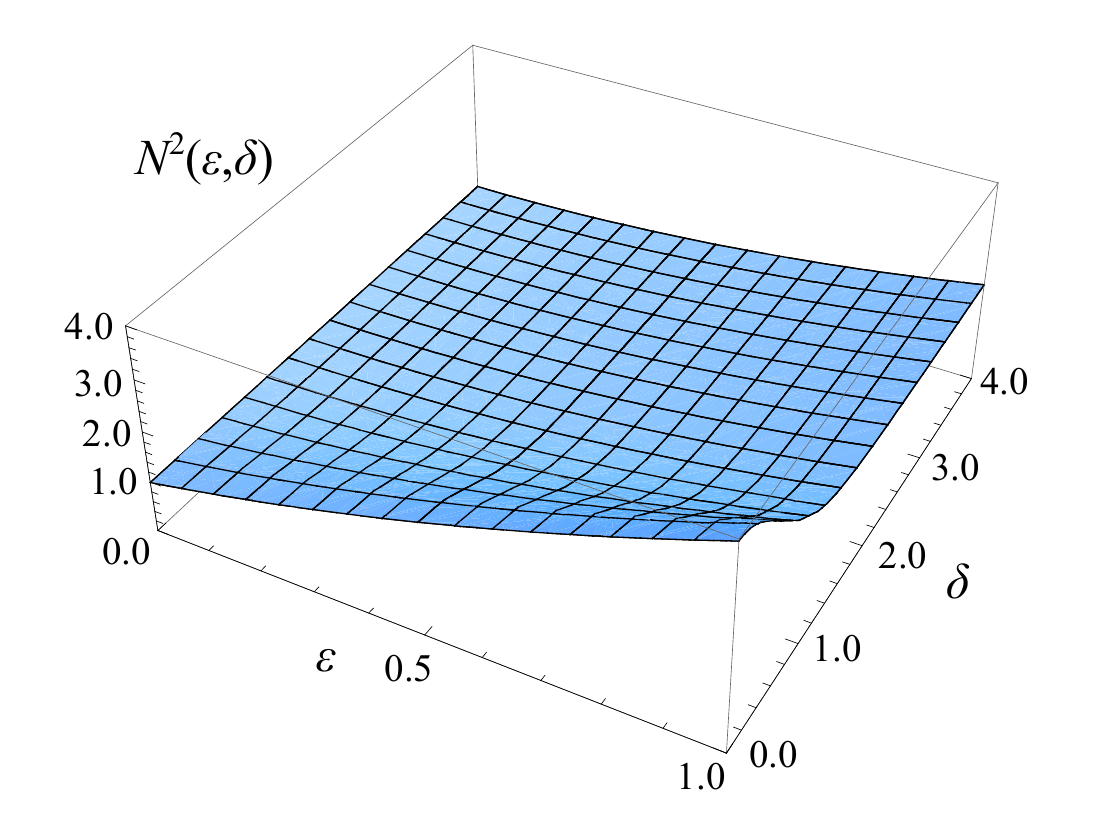}
\caption{Three-dimensional plot of $N^2(\epsilon,\delta)$.}
\label{fig:Nsqd}
\end{figure}
Figure \ref{fig:Nsqd} shows a plot of $N^2(\epsilon,\delta)$ for the  values of 
the $a$'s, $b$'s, and $\mu$'s that were fitted to the normalized $\psi_a$ and $\psi_b$, Eq.\ (\ref{abmus}).
We have checked that, for these values, $N^2(0,0) = 0.9858 \approx 1$  
and $N^2(1,0) = 3.9430 \approx 4$, as they should but with some 
deviation ($\approx 2$\%) coming from the inexactness of the fitting.
The ratio of the two values is 4 to high accuracy.

\begin{figure} 
\includegraphics[width=0.6\textwidth, height=0.5\textwidth, angle=0]{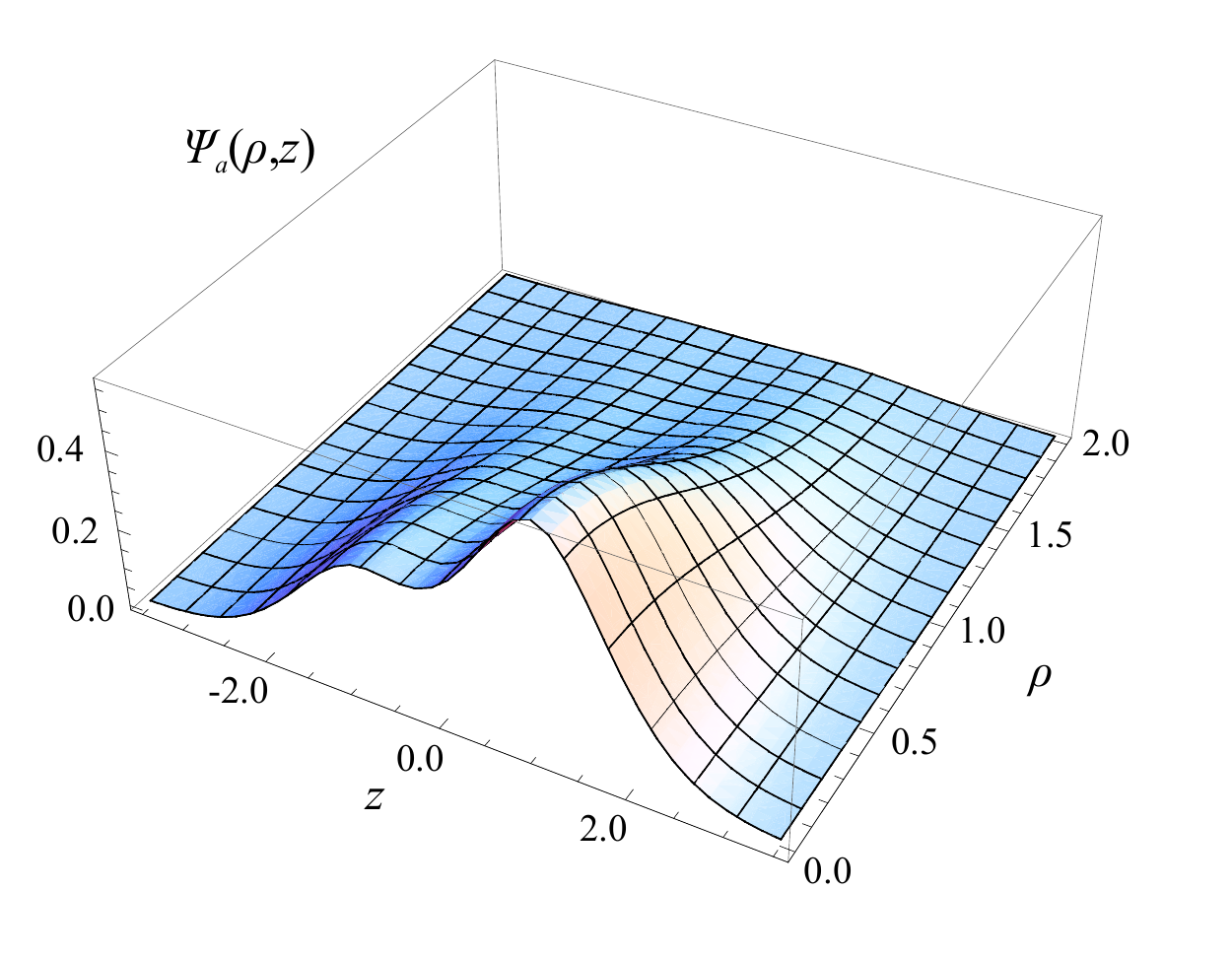}
\caption{Plot of a normalized $\Psi_a(\rho,z)$ for $\epsilon=0.5$ and $\delta = 1.0$.}
\label{fig:Psia}
\end{figure}
To illustrate what ''leakage'' from one well to the other might look like,
Fig.\ \ref{fig:Psia} shows a plot of the upper component of the normalized $\Psi$ 
as a function of $\rho$ (running from 0 to 2) and $z$ (running from -3.5 to +3.5) 
for $\epsilon = 0.5$ and $\delta = 1.1$.

\subsection{Evaluating the diagonal expectation $\left<\,- \nabla^2 \,\right>$}

First, note that, for ${\bf r}_\pm = \{x,\,y,\,z \pm \delta\}$, the $i$th component of the gradient 
\Be 
	\nabla_i = \frac{\partial}{\partial x_i} = \nabla_i' = \frac{\partial} {\partial x_i'}
	\quad \mbox{for} \quad {\bf r}' = \{x'=x,\,y'=y,\,z'=z\pm\delta\} = {\bf r}_\pm \,
\Ee
since each $\partial x_i'/\partial x_i = 1$.  Thus we can replace the result of the 
Laplacian with respect to $r$ acting on a function such as $\psi_a(r_-)$ 
with that for a Laplacian with respect to $r_-$ acting on that function.
For spherical coordinates, $-\nabla^2$ on the angle-independent $e^{-\mu_j\, r_-^2/2}$ then becomes
\Be 
	-\nabla^2 \;e^{-\mu_j r_-^2/2} = -\nabla'^{\,2} \;e^{-\mu_j\, r_-^2/2} = 
		-\frac{1}{r_-} \frac{d^2}{d\, r_-^2} \left(\; r_- e^{-\mu_j\, r_-^2/2} \;\right) =
		-\mu_j(\mu_j\, r_-^2 - 3)\; e^{-\mu_j\, r_-^2/2} \ , \label{minusdelsqGauss}
\Ee
whence the three-dimensional integral reduces, after symmetrizing and cancelling factors of 
$4\pi$, to
\Be
	I_{ij,\;<-\nabla^2>}^{(0)} 
		= 3\; \mu_i\mu_j \left[\frac{\pi}{2(\mu_i + \mu_j)^5}\right]^{1/2} , \label{LapI0}
\Ee
independent of $\delta$.

For the $B$-integrals things are more complicated because of the  
$\mbox{\boldmath $\sigma$} \cdot {\bf r}_-$ factor to the right of the Laplacian.
After some algebra,
\Be
	-({\mbox{\boldmath $\sigma$} \cdot {\bf r}_-} \nabla^2 
	{\mbox{\boldmath $\sigma$} \cdot {\bf r}_-}) \,e^{- \mu_j r_-^2/2}  
	= - r_-^2 \; \mu_j\,(\mu_j r_-^2 - 5) \,e^{- \mu_j r_-^2/2} \label{B0dots}
\Ee
whence
\Be
	J_{ij,\;<-\nabla^2>}^{(0)} 
		= \;15 \;\mu_i\mu_j \left[\frac{\pi}{2(\mu_i + \mu_j)^7}\right]^{1/2} \ ,
\Ee
also independent of $\delta$.
	 
The cross terms, again, do depend on $\delta$.
\Be
	I_{ij,\;<-\nabla^2>}^{(1)} =  \mu_i\mu_j\, [\,3\,(\mu_i+\mu_j) - 4\mu_i\mu_j\,\delta^2\,]
			\left[ \frac{2\pi}{(\mu_i+\mu_j)^7} \right]^{1/2} 
			\; e^{-2\mu_i\mu_j\,\delta^2/(\mu_i+\mu_j)} \ .
\Ee
This integral as a function of $\delta$ looks like Fig.\ \ref{sixplots}B.

For the corresponding $B$-cross term, one proceeds in the same manner but, instead of
Eq.\ (\ref{B0dots}), we need\footnote{
Because we have separated the two wells along the $z$-direction, the cross product 
${\bf r}_+ \times {\bf r_-}$ only has $x$ and $y$ components.
Since we have assumed the Pauli spinor $\chi_{m_s}$ to be polarized along the $z$-axis, 
the term from the product of two Pauli $\sigma$ matrices that gives a 
$i \mbox{\boldmath $\sigma$} \cdot {\bf r}_+ \times {\bf r_-}$ contribution vanishes.}
\Be
	-({\mbox{\boldmath $\sigma$} \cdot {\bf r}_+} \nabla^2 
	{\mbox{\boldmath $\sigma$} \cdot {\bf r}_-}) \,e^{- \mu_j r_-^2/2}  
	= -(\rho^2 + z^2 - \delta^2)  \; \mu_j\,(\mu_j r_-^2 - 5) \,e^{- \mu_j r_-^2/2} \ .
\Ee 
We find
\Bea
	J_{ij,\;<-\nabla^2>}^{(1)} &=&  \mu_i\mu_j\,
			[\,15\,(\mu_i+\mu_j)^2 - 40\,\mu_i\mu_j (\mu_i+\mu_j) \,\delta^2 + 16\,\mu_i^2\mu_j^2\;\delta^4 \,] 	
				\times \nonumber \\
		& & \qquad \left[ \frac{2\pi}{(\mu_i+\mu_j)^{11}} \right]^{1/2} 
				 e^{-2\mu_i\mu_j\,\delta^2/(\mu_i+\mu_j)}  \ .
\Eea
This integral as a function of $\delta$ also looks like Fig.\ \ref{sixplots}B,
but because it is quartic, it is slightly positive beyond $\delta = 1.7$.

\subsection{ Evaluating the expectation of $V^2+S^2+2\beta \,VS$}

This is also a diagonal operator.
The linear vector potential $V({\bf r})$ differs from the linear scalar potential $S({\bf r})$ by a
negative offset $-\;R$.
In $\left<V^2({\bf r})\right>$ the integrals of $\left< r_\pm^2 \right>$ are the same as those 
for $\left< S^2({\bf r}) \right>$.
Here, $\left< r_\pm^2 \right>$ means the integration of $r_-^2$ when $z > 0$ and of
$r_+^2$ when $z < 0$.
Thus we (schematically) expand the diagonal $V^2 + S^2 + 2\beta\,VS$ as
\Be
	\left< V^2 + S^2 + 2\beta\,VS \right> = 2 \left<\; r_\pm^2 \;\right> (1 + \beta) -
		 2\;R \left< \;r_\pm \;\right>(1 + \beta) + \;R^2 \left< \;1\; \right>  \ . \label{V2S22VS}
\Ee
The factor of $(1 + \beta)$ ensures that only the upper components of $\Psi$ contribute to the
first two expectation values.
That is, we only need to calculate the $A$-integrals (the $I$'s) for those terms.
The expectation value $\left< \,1\, \right>$ multiplying $R^2$ does have contributions 
from the lower components and their integrals $I_{<1>}^{(0)}$, $I_{<1>}^{(1)}$, 
$J_{<1>}^{(0)}$, and $J_{<1>}^{(1)}$ are given in Sec.\ \ref{normPsi}..
The integrals for the operators $\left< r_\pm^2 \right>$ and $\left< r_\pm \right>$ are 
rather more complicated and their analytic forms are presented next.

\subsubsection{Expectation of ${\cal O}_D = r_\pm^2$}

The direct integral for this operator is
\Bea
	I_{ij,\;<r_\pm^2>}^{(0)} &=& -\;\frac{2\delta}{(\mu_i + \mu_j)^2 } e^{-(\mu_i + \mu_j)\;\delta^2/2} \nonumber \\
		 & & + \;\left[ \frac{\pi}{2(\mu_i + \mu_j)^5} \right]^{1/2} \!
			\left[ 3 + 2 (\mu_i + \mu_j)\;\delta^2 \;
			\text{Erfc}\left (\sqrt{\frac{(\mu_i + \mu_j)}{2}}\;\delta \right) \right] \ .
			\label{Irpmsqd0}
\Eea
Note the linear dependence on $\delta$,
which gives rise to a shallow minimum near the origin before the function
returns to its initial value, as in Fig. \ref{sixplots}C.

The cross-term integral for $<r_\pm^2>$ is
\Bea
	I_{ij,\;<r_\pm^2>}^{(1)} & =& \;-\;\frac{4\delta}{(\mu_i + \mu_j)^2 } 
				e^{-(\mu_i + \mu_j)\;\delta^2/2}   \nonumber \\
			 & & \;+ \; \left[ \frac{2\pi} {(\mu_i + \mu_j)^7} \right]^{1/2} 
			 	e^{-2\mu_i\mu_j\;\delta^2/(\mu_i + \mu_j)} \times  \\
			 & & \quad\quad \left\{
				3(\mu_i + \mu_j) + 2\;(\mu_i^2+\mu_j^2)\;\delta^2  
				- 2\; (\mu_i^2 - \mu_j^2)\;\delta^2 \; 
				\text{Erf}\left( \frac{(\mu_i - \mu_j)} {\sqrt{2(\mu_i + \mu_j)}}\;\delta \right)
				\right\}, \nonumber \label{Irpmsqd1}
\Eea
which also has odd terms in $\delta$.
In this case, as a function of $\delta$, $I_{ij,\;<r_\pm^2>}^{(1)}$ falls off smoothly to zero from its
peak value at $\delta = 0$, as in Fig. \ref{sixplots}D.
$I_{ij,\;<r_\pm^2>}^{(1)}$ is symmetric in $i$ and $j$ because 
$\text{Erf}(-x) = -\text{Erf}(x)$, $I_{ij,\;<r_\pm^2>}^{(1)} = I_{ji,\;<r_\pm^2>}^{(1)}$. 
Also, as expected,  
\Be
	I_{<r_\pm^2>}^{(1)} = \;2\;I_{<r_\pm^2>}^{(0)} = 
		3 \left[\frac{2\pi} {(\mu_i + \mu_j)^5}\right]^{1/2} \ 
\Ee
when $\delta = 0$.

\subsubsection{Expectation of ${\cal O}_D = r_\pm $ \label{exprpm}}

The direct term for this operator is
\Bea
	I_{ij,\;<r_\pm >}^{(0)} & =& \; \frac{1}{2(\mu_i + \mu_j)^2} \;
			[4 + 2\;e^{-2(\mu_i + \mu_j)\;\delta^2} - 3\; e^{-(\mu_i + \mu_j)\;\delta^2/2}]  
				\nonumber \\
		& & - \ \frac{1}{2\delta}\; \left[\frac{\pi}{2(\mu_i + \mu_j)^5}\right]^{1/2} \times  \\
		& & \qquad\qquad	\left\{ \;(\mu_i + \mu_j)\,\delta^2    
			- \ \left(1 + 4(\mu_i + \mu_j)\;\delta^2\right)\;
			\text{Erf}\left( \sqrt{2\,(\mu_i + \mu_j)}\ \delta \right)  \right.  \nonumber \\ 
		& & \left. 	\qquad\qquad\qquad \ + \ 
			\left( 1 + 3(\mu_i + \mu_j)\;\delta^2 \right)\;
			\text{Erf}\left( \sqrt{(\mu_i + \mu_j)/2}\ \delta \, \right) \right\} \ .  \nonumber
\Eea
This integral also has an odd term in $\delta$, like $I_{ij,\;<r_\pm^2 >}^{(0)}$.
As a function of $\delta$ it resembles that shown in Fig.\ \ref{sixplots}C.
That is, despite the $1/\delta$ factor in the last term, $I_{ij,\;<r_\pm >}^{(0)}$ is {\it not} singular
at $\delta = 0$ (i.e., when there is no separation between the two wells):  
$I_{ij,\;<r_\pm >}^{(0)} \rightarrow 2/(\mu_i + \mu_j)^2$ as $\delta \rightarrow 0\ $.

The cross term for $\left< r_\pm \right>$ is 
\Bea
	I_{ij,\;<r_\pm >}^{(1)} &=& \; \frac{2} {(\mu_i+\mu_j)^2}  
				\ \left(e^{-2\mu_i\;\delta ^2} + e^{-2\mu_j\;\delta ^2} 
					- e^{-\frac{1}{2}(\mu_i+\mu_j)\;\delta ^2} \right) \nonumber \\ 
		& & + \ \frac{1}{2\, \delta \, \mu_i \mu_j} \left[ \frac{\pi}{2(\mu_i+\mu_j)^5} \right]^{1/2}  	
		\left\{\left(\mu_i+\mu_j\right)^2 \;\text{Erfc}\left(\sqrt{\frac{\mu_i+\mu_j}{2}}\,\delta \right)  
			\right. \nonumber \\ & &\quad\quad \left.
		-\;2\;\mu_j \left(\mu_i+\mu_j + 4\,\mu_i^2\,\delta ^2  \right)\;
		e^{-\frac{2 \delta ^2 \mu_i \mu_j}{\mu_i+\mu_j}}\;
		\text{Erfc}\left( \sqrt{\frac{2}{\mu_i+\mu_j}} \;\mu_i\,\delta  \right)  
			\right. \nonumber \\ & &\quad\quad \left.
		+\;2\text{  }\mu_i \left(\mu_i+\mu_j + 4\,\mu_j^2\,\delta ^2 \right)\;
		e^{-\frac{2 \delta ^2 \mu_i \mu_j}{\mu_i+\mu_j}} \;
		\text{Erf}\left( \sqrt{\frac{2}{\mu_i+\mu_j}} \;\mu_j\,\delta  \right)  
			\right. \nonumber \\ & &\quad\quad \left.
		-\;\left(\mu_i-\mu_j\right) \left( \mu_i+\mu_j - 4\;\mu_i\mu_j\,\delta ^2 \right)
		e^{-\frac{2 \delta ^2 \mu_i \mu_j}{\mu_i+\mu_j}}\;
		\text{Erfc}\left(\frac{ \left(\mu_i-\mu_j\right)\,\delta }{\sqrt{2\,(\mu_i+\mu_j)}}\right) 
			\right\}
\Eea
Note that $I_{ij,\;<r_\pm >}^{(1)}$ is also symmetric under the interchange of $i$ and 
$j$ and, again, at $\delta = 0$, we have
$I_{ij,\;<r_\pm >}^{(1)} = 4/(\mu_i + \mu_j)^2 = 2\;I_{ij,\;<r_\pm >}^{(0)}$.
Its behavior as a function of $\delta$ is similar to that shown in Fig.\ \ref{sixplots}D,
again partly due to the presence of odd terms in $\delta$.

\subsection{ The off-diagonal expectation of 
$-i\mbox{\boldmath $\alpha$} \cdot \left[ \left({\bf \nabla}
V({\bf r})\right) +  \beta\left({\bf \nabla} S({\bf r})\right) \right]$}

For the linear potentials of Eq.\ (\ref{VStwowell})
\Be
	 {\bf \nabla} V({\bf r}) \;=\; {\bf \nabla} S({\bf r}) = 
		\left\{ \begin{array}{ll}
			{\bf \hat{r}_-} & \ \mbox{if $z > 0$} \\
			{\bf \hat{r}_+} & \ \mbox{if $z < 0$} 
				\end{array} 
				\right.		\  
\Ee
and we again have a simplification from the $(1 + \beta)$, namely,
\Be
	-i\mbox{\boldmath $\alpha$} \cdot \left[ \left({\bf \nabla} V({\bf r})\right) 
			+  \beta\left({\bf \nabla} S({\bf r})\right) \right] =
	-i\mbox{\boldmath $\alpha$} \cdot {\bf \hat{r}_\pm} (1+\beta) = 
		\left[ \begin{array}{cc}
		                 0                                  & \quad 0 \\
		-2i\;\mbox{\boldmath $\sigma$} \cdot {\bf \hat{r}_\pm} & \quad 0 
						\end{array} 
		\ \right] \ ,
\Ee
i.e., the operator ${\bf X}_{12}$ in Eq.\ (\ref{Xij}) vanishes and ${\bf X}_{21}$ is doubled.
The latter operator connects the upper component of $\Psi^\dagger$ to the lower component of $\Psi$.

For the direct terms, Eq.\ (\ref{formOOD0}) reduces to two terms 
\Bea
	<[\nabla VS]^{(0)}> &=& -2\; \frac{1}{4\pi}\; \sum_{i,j} \int d^3 r \; \left\{ 
			e^{-\mu_i \,r^2_-/2}\; \left[ a_j b_i \,
			(\mbox{\boldmath $\sigma$} \cdot {\bf r}_-) 
			(\mbox{\boldmath $\sigma$} \cdot {\bf \hat{r}}_\pm) 
			\right] \; e^{-\mu_j \,r^2_-/2} \right. \nonumber \\
		& & \qquad\qquad\quad \left.  +\;
			e^{-\mu_j \,r^2_-/2}\; \left[ a_i b_j \,
			(\mbox{\boldmath $\sigma$} \cdot {\bf r}_-)  
			(\mbox{\boldmath $\sigma$} \cdot {\bf \hat{r}}_\pm) 
			\right] \; e^{-\mu_i \,r^2_-/2}    
			\right\}  \nonumber \\
		&=& \sum_{i,j} \left[ a_j b_i \,K_{ij,\,<\nabla VS>}^{(0)} + 
			a_i b_j \,K_{ji,\,<\nabla VS>}^{(0)} \right] \ ,
\Eea
where  
\Be
	K_{ij,<\nabla VS>}^{(0)} = -2\; \frac{1}{4\pi}\; \int d^3 r \;  
			e^{-\mu_i \,r^2_-/2}\; \left[\,
			(\mbox{\boldmath $\sigma$} \cdot {\bf r}_-) 
			(\mbox{\boldmath $\sigma$} \cdot {\bf \hat{r}}_\pm) 
			\right] \; e^{-\mu_j \,r^2_-/2} \ .
\Ee
The Pauli matrices here reduce to
\Be
	(\mbox{\boldmath $\sigma$} \cdot {\bf r}_-) \;
		(\mbox{\boldmath $\sigma$} \cdot {\bf \hat{r}_\pm}) = 
		r_- \;({\bf \hat{r}_-} \cdot {\bf \hat{r}_\pm}) \ . \label{sigsK0}
\Ee
For the integration over $z>0$ the integrand becomes simply $r_-$,
which is the same as that already needed for getting to the final result 
for $I_{<r_\pm >}^{(0)}$ in subsection \ref{exprpm} above.
For the integration over negative $z$, however, Eq.\ (\ref{sigsK0}) becomes
\Be
	r_- \;({\bf \hat{r}_-} \cdot {\bf \hat{r}_+}) = {\bf r}_- \cdot {\bf r}_+ / r_+ 
	= (\rho^2 + z^2 - \delta^2)/\sqrt{\rho^2 + (z + \delta)^2} \ ,
\Ee
which involves a new integrand, but which nonetheless can still be done analytically.
(Here it is much easier to do the $\rho$-integration first.)
We find 
\Bea
	\!\!\!\!\!\! K_{ij,<\nabla VS>}^{(0)} 
		&=& -\ \frac{2}{(\mu_i + \mu_j)^2} 
			\left[ 2 - e^{-(\mu_i + \mu_j)\;\delta^2/2} \right]  \nonumber \\
	 	& &	- \ \frac{1}{\delta}\, \left[\frac{2\pi}{(\mu_i + \mu_j)^5}\right]^{1/2}
	 		\left[ \text{Erf}\left( \sqrt{2(\mu_i + \mu_j)}\;\delta \,\right) 
	 		 	-  \text{Erf}\left( \sqrt{(\mu_i + \mu_j)/2}\;\delta \,\right) \right]\ .  
\Eea
This result is, again, symmetric and non-singular with 
$K_{ij,<\nabla VS>}^{(0)} = -4/(\mu_i + \mu_j)^2$ at $\delta = 0$.
In this case there are {\it no} odd terms (!) in $\delta$.
Versus $\delta$ it is similar to that shown in Fig.\ \ref{sixplots}C,
but with the initial slope at the origin being zero.

Because $K_{ij,<\nabla VS>}^{(0)} = K_{ji,<\nabla VS>}^{(0)}$, we can finally write the direct term
contributions for this expectation as
\Be
	<[\nabla VS]^{(0)}> \ = \sum_{i,j} \left( a_j b_i + a_i b_j \right)\; \,K_{ij,\,<\nabla VS>}^{(0)} 
		\label{nablaVS0} \ ,
\Ee
regaining explicit symmetry.

The cross term integral $K_{<\nabla VS>}^{(1)}$ is more complicated but is done similarly.
As ${\bf X}_{12} = 0$, there are now four terms remaining from Eq.\ (\ref{formOOD1}),
\Bea
	<[\nabla VS]^{(1)}> &=& -2\; \frac{1}{4\pi}\; \sum_{i,j} \int d^3 r \; \left\{ 
		e^{-\mu_i \,r^2_+/2}\; \left[ a_j b_i \,
			(\mbox{\boldmath $\sigma$} \cdot {\bf r}_+) 
			(\mbox{\boldmath $\sigma$} \cdot {\bf \hat{r}}_\pm) 
			\right] \; e^{-\mu_j \,r^2_-/2} \right. \nonumber \\
		& & \qquad\qquad \left.  +\;
			e^{-\mu_i \,r^2_-/2}\; \left[ a_j b_i \,
			(\mbox{\boldmath $\sigma$} \cdot {\bf r}_-)  
			(\mbox{\boldmath $\sigma$} \cdot {\bf \hat{r}}_\pm) 
			\right] \; e^{-\mu_j \,r^2_+/2} \right. \nonumber \\
		& &	\qquad\qquad \left.  +\;
			e^{-\mu_j \,r^2_+/2}\; \left[ a_i b_j \,
			(\mbox{\boldmath $\sigma$} \cdot {\bf r}_+) 
			(\mbox{\boldmath $\sigma$} \cdot {\bf \hat{r}}_\pm) 
			\right] \; e^{-\mu_i \,r^2_-/2} \right. \nonumber \\
		& & \qquad\qquad \left.  +\;
			e^{-\mu_j \,r^2_-/2}\; \left[ a_i b_j \,
			(\mbox{\boldmath $\sigma$} \cdot {\bf r}_-)  
			(\mbox{\boldmath $\sigma$} \cdot {\bf \hat{r}}_\pm) 
			\right] \; e^{-\mu_i \,r^2_+/2}    
			\right\}  \nonumber \\
		&=& \sum_{i,j} \int d^3 r \; \left[ a_j b_i \; K_{ij,\,<\nabla VS>}^{(1)} +
			a_i b_j \; K_{ji,\,<\nabla VS>}^{(1)} \right] \ ,
\Eea
where 
\Bea
	K_{ij,\,<\nabla VS>}^{(1)} &=& -2\; \frac{1}{4\pi}\; \int d^3 r \; \left\{
		e^{-\mu_i \,r^2_+/2}\; \left[  \,
			(\mbox{\boldmath $\sigma$} \cdot {\bf r}_+) 
			(\mbox{\boldmath $\sigma$} \cdot {\bf \hat{r}}_\pm) 
			\right] \; e^{-\mu_j \,r^2_-/2} \right. \nonumber \\
		& & \qquad\qquad \left.  +\;
			e^{-\mu_i \,r^2_-/2}\; \left[  \,
			(\mbox{\boldmath $\sigma$} \cdot {\bf r}_-)  
			(\mbox{\boldmath $\sigma$} \cdot {\bf \hat{r}}_\pm) 
			\right] \; e^{-\mu_j \,r^2_+/2} \,\right\} 
\Eea
In addition to Eq.\ (\ref{sigsK0}) we also need
\Be
	(\mbox{\boldmath $\sigma$} \cdot {\bf r}_+) \;
		(\mbox{\boldmath $\sigma$} \cdot {\bf \hat{r}_\pm}) = 
		r_+ \;({\bf \hat{r}_+} \cdot {\bf \hat{r}_\pm}) \ , \label{sigsK1}
\Ee
which becomes $r_+$ for the $z < 0$ integration and 
$(\rho^2 + z^2 - \delta^2)/\sqrt{\rho^2 + (z - \delta)^2}$ for the $z > 0$ integration.

The integrations over $z$ go much easier if one re-defines the integrations over $z$
in terms of $\mu = \mu_i + \mu_j$ and $\nu = \mu_i - \mu_j$.  
The resulting integrals in $\mu$ and $\nu$ can then be converted back to $\mu_i$ and $\mu_j$.
We find
\Bea
	K_{ij,\,<\nabla VS>}^{(1)} &=&  
		\ \frac{1}{\mu_i\mu_j (\mu_i + \mu_j)^2} \;  
		\left[\; 2\,\mu_j\,(\mu_j - \mu_i)\;e^{-2 \mu_i \;\delta^2} + 
			2\,\mu_i\,(\mu_i - \mu_j)\;e^{-2 \mu_j \;\delta^2} \right. \nonumber \\
		& & \left.\qquad\qquad\qquad\qquad\qquad  -\;(\mu_i - \mu_j)^2\;e^{-(\mu_i +\mu_j) \;\delta^2/2}
		 	\;\right]  \nonumber \\
	& &	- \ \frac{1} {2\,\delta\;\mu_i^2\mu_j^2}  \left[ \frac{\pi} {2 (\mu_i + \mu_j)^5} \right]^{1/2}
				\times \\
	& &	\qquad\qquad	\left\{
			(\mu_i +\mu_j)^3\,\left(\, \mu_i + \mu_j - 2 \mu_i\mu_j\;\delta^2 \,\right) \right. 
				\;\text{Erfc}\left( \sqrt{(\mu_i + \mu_j)/2}\;\delta \,\right) 
			 \nonumber \\
	& & \qquad\qquad\quad + \ 2\,\mu_i^2\;
		\left[\; (\mu_i^2 + 4\mu_i\mu_j + 3\mu_j^2) - 4 \mu_j^2 (\mu_i - \mu_j)\;\delta^2 \;\right] 
			\  \nonumber \\
		& & \qquad\qquad\qquad\qquad\qquad \times\; e^{-2\mu_i\mu_j\;\delta^2/(\mu_i + \mu_j)}
			\;\text{Erf}\left( \sqrt{2/(\mu_i + \mu_j)}\;\mu_j\;\delta \,\right) \nonumber \\
	& & \qquad\qquad\quad - \  2\,\mu_j^2\; 
		\left[\; 3\mu_i^2 + 4\mu_i\mu_j + \mu_j^2 - 4 \mu_i^2 (\mu_j - \mu_i)\;\delta^2 \;\right] 
		\  \nonumber \\
		& & \qquad\qquad\qquad\qquad\qquad \times\; e^{-2\mu_i\mu_j\;\delta^2/(\mu_i + \mu_j)}
		\;\text{Erfc}\left( \sqrt{2/(\mu_i + \mu_j)}\;\mu_i\;\delta \,\right) \nonumber \\
	& & \qquad\qquad\quad - \ 
		\left[\; (\mu_i^3 + 5\,\mu_i^2\mu_j + 5\,\mu_i\mu_j^2 + \mu_j^3) - 
			8\,\mu_i^2\mu_j^2\;\delta^2 \;\right] \  \nonumber \\
		& & \qquad\qquad\qquad\quad \left.
			\times\,(\mu_i - \mu_j)\,e^{-2\mu_i\mu_j)\;\delta^2/(\mu_i + \mu_j)}
			\;\text{Erfc}\left( \frac{(\mu_i - \mu_j)\;\delta} {\sqrt{2(\mu_i + \mu_j)}} \,\right) 
			\right\} \nonumber \ , 
\Eea
which also is symmetric and goes to $-8/(\mu_i + \mu_j)^2 = 2\, K_{ij,\,<\nabla VS>}^{(0)}$ at $\delta = 0$.
This integral does have some odd terms in $\delta$.
As a function of $\delta$ it resembles a Gaussian, i.e., 
looks like that shown in Fig.\ \ref{sixplots}A.

Because $K_{ij,\,<\nabla VS>}^{(1)} = K_{ji,\,<\nabla VS>}^{(1)}$ we can again finally write
\Be
	<[\nabla VS]^{(1)}> \ = \sum_{i,j} \left( a_j b_i + a_i b_j \right)\; \,K_{ij,\,<\nabla VS>}^{(1)} 
		\label{nablaVS1} \ ,
\Ee
mirroring the form of Eq.\ (\ref{nablaVS0}).

\subsection{The off-diagonal expectation $-2i\; V({\bf r}) \;\mbox{\boldmath $\alpha$} \cdot \nabla$}

For this off-diagonal operator ${\bf X}_{12} = {\bf X}_{21} = -2\,V({\bf r})\,\nabla$ in Eq.\ (\ref{Xij})
and the direct term expectation, Eq.\ (\ref{formOOD0}), has all four terms
\Bea
	\!\!\!\!\!\!<[2\,V\nabla]^{(0)}> &=& \frac{1}{8\pi}\; \sum_{i,j} \int d^3 r \, V({\bf r}) \;\left\{ 
			e^{-\mu_i \,r^2_-/2}\; \left[ \;2\,a_i b_j 
			(\mbox{\boldmath $\sigma$} \cdot \nabla)
			(\mbox{\boldmath $\sigma$} \cdot {\bf r}_-) \right. \right. \nonumber \\
		& & \qquad\qquad\qquad\qquad\qquad\qquad \left. -\; 2\, b_i a_j  
			(\mbox{\boldmath $\sigma$} \cdot {\bf r}_-) 
			(\mbox{\boldmath $\sigma$} \cdot \nabla)\; 
			\right] \; e^{-\mu_j \,r^2_-/2}  \nonumber \\
		& & \qquad\qquad\qquad \left.  +\;
			e^{-\mu_j \,r^2_-/2}\; \left[ \;2\,a_j b_i 
			(\mbox{\boldmath $\sigma$} \cdot \nabla) 
			(\mbox{\boldmath $\sigma$} \cdot {\bf r}_-) \right. \right. \nonumber \\
		& & \qquad\qquad\qquad\qquad\qquad\qquad \left. \left. -\; 2\, b_j  a_i 
			(\mbox{\boldmath $\sigma$} \cdot {\bf r}_-)  
			(\mbox{\boldmath $\sigma$} \cdot \nabla) \;
			\right] \; e^{-\mu_i \,r^2_-/2}    
			\right\} \, . \label{Vnabla0}
\Eea
With 
\Be
	\nabla_k \; e^{-\mu_i\,r_-^2/2} = -\mu_i ({\bf r}_-)_k \; e^{-\mu_i\,r_-^2/2} , \quad
	\nabla_k ({\bf r}_-)_l = \delta_{kl} , \quad \text{ and } \quad
	(\mbox{\boldmath $\sigma$} \cdot \nabla) (\mbox{\boldmath $\sigma$} \cdot {\bf r}_-) = 3 \,
\Ee
we have, for the first terms in the square brackets of Eq.\ (\ref{Vnabla0}),
\Bea
	(\mbox{\boldmath $\sigma$} \cdot \nabla) && \!\!\!\!\!\!\!
		     	(\mbox{\boldmath $\sigma$} \cdot {\bf r}_-) \; e^{-\mu_i\,r_-^2/2} = 
	 \; e^{-\mu_i\,r_-^2/2} \; (\mbox{\boldmath $\sigma$} \cdot \nabla) 
		 	(\mbox{\boldmath $\sigma$} \cdot {\bf r}_-) +
		 \mbox{\boldmath $\sigma$}  \cdot \;[ (\mbox{\boldmath $\sigma$} \cdot {\bf r}_-) \nabla  
		 	\; e^{-\mu_i\,r_-^2/2} ] \nonumber\\
		 &=& [3 - \mu_i (\mbox{\boldmath $\sigma$} \cdot {\bf r}_-) 
		 	(\mbox{\boldmath $\sigma$} \cdot {\bf r}_-)]\; e^{-\mu_i\,r_-^2/2} 
		 	= \,(3 -\mu_i\, r_-^2 ) \  e^{-\mu_i\,r_-^2/2} \  \label{1stterm}
\Eea
and similarly when acting on $e^{-\mu_j\,r_-^2/2}$.

For the second terms in the square brackets of Eq.\ (\ref{Vnabla0}), 
\Bea
	(\mbox{\boldmath $\sigma$} \cdot {\bf r}_-) (\mbox{\boldmath $\sigma$} \cdot \nabla) \; e^{-\mu_i\,r_-^2/2}
		&=& -\mu_i (\mbox{\boldmath $\sigma$} \cdot {\bf r}_-) (\mbox{\boldmath $\sigma$} \cdot {\bf r}_-) 
			\; e^{-\mu_i\,r_-^2/2} 
		= -\mu_i r_-^2 \; e^{-\mu_i\,r_-^2/2} \ \label{2ndterm}
\Eea
and, again, similarly when acting on $e^{-\mu_j\,r_-^2/2}$. 
 
With Eqs.\ (\ref{1stterm}) and (\ref{2ndterm}), Eq.\ (\ref{Vnabla0}) reduces to
\Bea
	\!\!\!\!\!\!\!<[2\,V\nabla]^{(0)}> &=& \frac{1}{4\pi}\; \sum_{i,j} \int d^3 r \; 
		\left\{ e^{-\mu_i \,r^2_-/2}\; V(r_\pm) 
				\left[ a_i b_j\;(3 - \mu_j r_-^2)\;	+ \; b_i a_j \;  \mu_j r_-^2 \right] 
				\;e^{-\mu_j \,r^2_-/2} \right. \nonumber \\
			& & \qquad\qquad \left. +\; e^{-\mu_j \,r^2_-/2}\; V(r_\pm) 
				\left[ a_j b_i\;(3 -\mu_i r_-^2)\;	+ \; b_j a_i \; \mu_i r_-^2 \right] 
				\;e^{-\mu_j \,r^2_-/2} \right\}  \nonumber \\
		&=& \sum_{i,j} \left\{ a_i b_j \;K_{ij,\,<2V\nabla>}^{(0)} + 
			a_j b_i \;K_{ji,\,<2V\nabla>}^{(0)} \right\} 
\Eea
where, with $V(r_\pm) = r_\pm - R,$
\Bea
	K_{ij,\,<2V\nabla>}^{(0)} &=& \frac{1}{4\pi}\; \int d^3 r \; e^{-\mu_i \,r^2_-/2}\; (r_\pm - R)\;
		[(\mu_i- \mu_j)\, r_-^2 - 3] \; e^{-\mu_j\,r_-^2/2} \ . \label{K2Vnabla0}  \nonumber \\ 
	&=& (\mu_i - \mu_j)\;K_{ij,\,a}^{(0)} 
		- (\mu_i - \mu_j)\;R\;K_{ij,\,b}^{(0)}
		+ 3\; K_{ij,\,c}^{(0)} - 3\;R\; K_{ij,\,d}^{(0)}
\Eea
where these four integrals are
\Bea
	K_{ij,\,a}^{(0)} &=& \frac{1}{4\pi}\; \int d^3 r \; 
			e^{-\mu_i \,r^2_-/2}\; r_\pm \, r_-^2 \; e^{-\mu_j\,r_-^2/2} \nonumber \\
		&=& \ \frac{1}{2\,(\mu_j + \mu_i)^3} \left[\;16 + 6\;e^{-2(\mu_j + \mu_i)\;\delta^2}
			-11\;e^{-(\mu_j + \mu_i)\;\delta^2/2} \;\right] \nonumber \\
		& & + \frac{1}{2\delta} \left[ \frac{\pi}{2(\mu_j + \mu_i)^7} \right]^{1/2} 
			\left\{\; [5 + 9 (\mu_j + \mu_i) \delta^2]\; 
			\text{Erfc} \left( \sqrt{(\mu_j + \mu_i)/2}\; \delta \right) \right. \nonumber \\
		& & \qquad\qquad\qquad\qquad\qquad \left. -\;[5 + 12 (\mu_j + \mu_i) \delta^2]\; 
			\text{Erfc} \left( \sqrt{2(\mu_j + \mu_i)}\; \delta \right)
			\right\} \ , \\
	K_{ij,\,b}^{(0)} &=& \frac{1}{4\pi}\; \int d^3 r \; 
			e^{-\mu_i \,r^2_-/2}\; r_-^2 \; e^{-\mu_j\,r_-^2/2} 
			= 3 \;\left[\frac{\pi} {2\,(\mu_j + \mu_i)^5}\right]^{1/2} \ , \\
	K_{ij,\,c}^{(0)} &=& \frac{1}{4\pi}\; \int d^3 r \; 
			e^{-\mu_i \,r^2_-/2}\; r_\pm \; e^{-\mu_j\,r_-^2/2} 
			\;=\; I_{ij,\;<r_\pm >}^{(0)} \ , \\
	K_{ij,\,d}^{(0)} &=& \frac{1}{4\pi}\; \int d^3 r \; e^{-\mu_i \,r^2_-/2} \; e^{-\mu_j\,r_-^2/2} 
			\;=\; I_{ij,\;<1>}^{(0)} \ .
\Eea
$K_{ij,\,a}^{(0)}$ has an odd term in $\delta$ and its plot resembles that shown in Fig.\ \ref{sixplots}D.
All four of the above integrals are symmetric in $i$ and $j$, so we can finally write
\Be
	<[2\,V\nabla]^{(0)}> \; = \;\sum_{i,j} \left( a_j b_i + a_i b_j \right)\; \,K_{ij,\,<2\,V\nabla>}^{(0)} 
		= 2\;\sum_{i,j}  a_j b_i \; K_{ij,\,<2\,V\nabla>}^{(0)} \ .
\Ee

For the cross term, from Eqs.\ (\ref{1stterm}) and (\ref{2ndterm}) and the like, 
Eq.\ (\ref{formOOD1}) becomes
\Bea
	& & <[2\,V\nabla]^{(1)}> \ = \ \frac{1}{8\pi}\; \sum_{i,j} \int d^3 r \; V(r_\pm) \ \times  \nonumber \\
		& &	\qquad\qquad \  \left\{\; 
			e^{-\mu_i \,r^2_+/2}\; \left[ \;2\,a_i b_j 
			(\mbox{\boldmath $\sigma$} \cdot \nabla)
			(\mbox{\boldmath $\sigma$} \cdot {\bf r}_-)  - \;2\, b_i a_j  
			(\mbox{\boldmath $\sigma$} \cdot {\bf r}_+) 
			(\mbox{\boldmath $\sigma$} \cdot \nabla) \;
			\right] \; e^{-\mu_j \,r^2_-/2}  \right. \nonumber \\
		& & \qquad\qquad\quad \left.  +\;
			e^{-\mu_i \,r^2_-/2}\; \left[ \;2\, a_i b_j 
			(\mbox{\boldmath $\sigma$} \cdot \nabla)
			(\mbox{\boldmath $\sigma$} \cdot {\bf r}_+)  - \;2\, b_i a_j 
			(\mbox{\boldmath $\sigma$} \cdot {\bf r}_-) 
			(\mbox{\boldmath $\sigma$} \cdot \nabla) \;
			\right] \; e^{-\mu_j \,r^2_+/2} \right. \nonumber \\
		& & \qquad\qquad\quad \left.  +\;
			e^{-\mu_j \,r^2_+/2}\; \left[ \;2\,a_j b_i 
			(\mbox{\boldmath $\sigma$} \cdot \nabla) 
			(\mbox{\boldmath $\sigma$} \cdot {\bf r}_-)  -\;2\, b_j a_i 
			(\mbox{\boldmath $\sigma$} \cdot {\bf r}_+)  
			(\mbox{\boldmath $\sigma$} \cdot \nabla) \;
			\right] \; e^{-\mu_i \,r^2_-/2}  \right. \nonumber \\
		& & \qquad\qquad\quad \left.  +\;
			e^{-\mu_j \,r^2_-/2}\; \left[ \;2\,a_j b_i 
			(\mbox{\boldmath $\sigma$} \cdot \nabla)
			(\mbox{\boldmath $\sigma$} \cdot {\bf r}_+)  - \;2\, b_j a_i 
			(\mbox{\boldmath $\sigma$} \cdot {\bf r}_-) 
			(\mbox{\boldmath $\sigma$} \cdot \nabla) \;
			\right] \; e^{-\mu_i \,r^2_+/2}   
			\;\right\} \nonumber \\
		& &  \qquad = \ \frac{1}{4\pi}\; \sum_{i,j}  \int d^3 r \; \; V(r_\pm)  \left\{\; 
			e^{-\mu_i \,r^2_+/2}\; \left[ a_i b_j 
			(3 - \mu_j r_-^2)   + \; b_i a_j \mu_j ({\bf r}_+ \cdot {\bf r}_-) \right] \; 
			e^{-\mu_j \,r^2_-/2}  \right. \nonumber \\
		& & \qquad\qquad\qquad\qquad \left.  +\;
			e^{-\mu_i \,r^2_-/2}\; \left[ a_i b_j 
			(3 - \mu_j r_+^2)  + \; b_i a_j \mu_j ({\bf r}_+ \cdot {\bf r}_-) \right] \; 
			e^{-\mu_j \,r^2_+/2} \right. \nonumber \\
		& & \qquad\qquad\qquad\qquad \left.  +\;
			e^{-\mu_j \,r^2_+/2}\; \left[ a_j b_i 
			(3 - \mu_i r_-^2)   + \; b_j a_i \mu_i ({\bf r}_+ \cdot {\bf r}_-) \right] \; 
			e^{-\mu_i \,r^2_-/2}  \right. \nonumber \\
		& & \qquad\qquad\qquad\qquad \left.  +\;
			e^{-\mu_j \,r^2_-/2}\; \left[ a_j b_i 
			(3 - \mu_i r_+^2)  + \; b_j a_i \mu_i ({\bf r}_+ \cdot {\bf r}_-) \right] \; 
			e^{-\mu_i \,r^2_+/2}
			\;\right\}  \nonumber \\ 
		& & \qquad = \ \sum_{i,j}  \left\{\; a_i b_j\;K_{ij,\,<2V\nabla>}^{(1)} 
			+ a_j b_i\;K_{ji,\,<2V\nabla>}^{(1)}
			\; \right\} \ , \label{Vnabla1}			
\Eea
where
\Bea
	K_{ij,\,<2\,V\nabla>}^{(1)} &=&  \frac{1}{4\pi}\; \int d^3 r \; (r_\pm - R) \left\{ \;
			e^{-\mu_i \,r^2_+/2}\; 
			\left[ (3 - \mu_j r_-^2) + \mu_i ({\bf r}_+ \cdot {\bf r}_-) \right]
			e^{-\mu_j \,r^2_-/2}\; \right. \nonumber \\
		& & \qquad\qquad\qquad\quad \left. +\;
			e^{-\mu_i \,r^2_-/2}\; 
			\left[ (3 - \mu_j r_+^2) + \mu_i ({\bf r}_+ \cdot {\bf r}_-) \right]
			e^{-\mu_j \,r^2_+/2}\; \right\} \nonumber \\
		&=& - \mu_j\,K_{ij,\,a}^{(1)} + \mu_j\,R\,K_{ij,\,b}^{(1)}  
			+ \mu_i\,K_{ij,\,c}^{(1)} - \mu_i\,R\,K_{ij,\,d}^{(1)}
			+ 3\,K_{ij,\,e}^{(1)} - 3\,R\,K_{ij,\,f}^{(1)}
				\ . \label{K2Vnabla1}
\Eea

The first integral,
\Be
	K_{ij,\,a}^{(1)} = \frac{1}{4\pi} \int d^3 r \;\left\{\; 
		e^{-\mu_i \,r^2_+/2}\; r_\pm \, r_-^2 \; e^{-\mu_j\,r_-^2/2} +
		e^{-\mu_i \,r^2_-/2}\; r_\pm \, r_+^2 \; e^{-\mu_j\,r_+^2/2} \right\} \ , \label{defK1a}
\Ee
can be done using $\mu = \mu_i + \mu_j$ and $\nu = \mu_i - \mu_j$, noting that $\mu > |\nu|$.  
Writing 
\Bea
	& & e^{-\mu_i \,r^2_+/2} \, r_-^2 \; e^{-\mu_j\,r_-^2/2} +
		e^{-\mu_i \,r^2_-/2} \, r_+^2 \; e^{-\mu_j\,r_+^2/2} \nonumber \\
	& & \qquad\qquad = 2\; e^{-\mu \,(\rho^2 + z^2 + \delta^2)/2} \left\{\; (\rho^2 + z^2 + \delta^2) \cosh (\nu \delta z)
				+ (2 z \delta) \sinh(\nu \delta z) \right\}	
\Eea
displays the $i,\;j$ symmetric and anti-symmetric parts explicitly.  
After converting back to $\mu_i$ and $\mu_j$,
\Bea
	 K_{ij,\,a}^{(1)} & & \;= \ \frac{2}{\mu_j(\mu_i+\mu_j)^4} 
		 \left\{\;[\;5\,\mu_j\, (\mu_i+\mu_j) + 4\, \mu_i^2 \mu_j \;\delta ^2\; ] 
				\;e^{-2\mu_i\;\delta ^2} \right. \nonumber \\
		& & \qquad\qquad\qquad\quad + \ 
				[\;(\mu_i+\mu_j)(-2 \mu_i  + 3 \mu_j) + 4\, \mu_i^2 \mu_j \;\delta ^2 \;]
		 		\;e^{-2  \mu_j\; \delta ^2} \nonumber \\
		& & \qquad\qquad\qquad\quad \left. + \ 
				[\,(\mu_i+\mu_j) (\mu_i - 4\,\mu_j)  - 4\, \mu_i^2\mu_j \;\delta ^2\,] 
		 		\;e^{-\frac{1}{2} (\mu_i+\mu_j)\; \delta ^2} \;\right\} \nonumber \\
		& & \nonumber \\
%
		& & + \ \frac{1}{2\,\delta\, \mu_i\, \mu_j^2}
			\left[ \frac{\pi}{2(\mu_i+\mu_j)^9} \right]^{1/2} \times  \\  
		& & \nonumber \\	
		& &  \left\{\;  
				 2\,[\;\mu_i\,(\mu_i+\mu_j)^2 (2\,\mu_i + 5\,\mu_j)
				+ \;4\,\mu_i\mu_j(\mu_i+\mu_j)
				(\mu_i^2 - 2\,\mu_i\mu_j + 3\,\mu_j^2)\;\delta^2 + 16\,\mu_i^3\mu_j^3\; \delta^4\;]
				 \right. \nonumber \\
		& & \qquad\qquad\qquad\qquad\qquad\qquad
				\times \;e^{-2\mu_i\mu_j\;\delta ^2/(\mu_i+\mu_j)} \;
				\text{Erf}\left( \sqrt{2/(\mu_i+\mu_j)}\;\mu_j\;\delta\;\right) \nonumber \\
		& & \quad - \ 
				2\,\mu_j^2[\;3\,(\mu_i + \mu_j)^2 + 24\,\mu_i^2 (\mu_i + \mu_j) \;\delta^2
				+ 16\,\mu_i^4\; \delta^4\;] \nonumber \\
		& & \qquad\qquad\qquad\qquad\qquad\qquad
				\times \;e^{-2\mu_i\mu_j\;\delta ^2/(\mu_i+\mu_j)} \;
				\text{Erfc}\left( \sqrt{2/(\mu_i+\mu_j)}\;\mu_i\;\delta\;\right) \nonumber \\
		& & \quad  - \ [\;(\mu_i+\mu_j)^2 (2\,\mu_i^2 + 5\,\mu_i\mu_j - 3\,\mu_j^2) 	
				+ \;4\,\mu_i\mu_j (\mu_i+\mu_j) (\mu_i^2 - 8\,\mu_i\mu_j + 3\,\mu_j^2)\;\delta^2
					 \nonumber \\
		& &	\qquad\qquad\qquad		-\; 16\;\mu_i^3\mu_j^2\,(\mu_i - \mu_j)\;\delta^4] 
				\;e^{-2\mu_i\mu_j\;\delta ^2/(\mu_i+\mu_j)} \;
				\;\text{Erfc}\left( \frac{(\mu_i-\mu_j)} {\sqrt{2(\mu_i+\mu_j)}} 
					\;\delta \;\right) \nonumber \\
		& & \quad  + \; \left.  (\mu_i+\mu_j)^3\,(2\,\mu_i + 3\,\mu_j )
				  \;\text{Erfc}\left( \sqrt{(\mu_i+\mu_j)/2}\;\delta\;\right) \right\}  \nonumber \ ,
\Eea
which is, as expected, not symmetric in $i$ and $j$.
It is, however, non-singular: $K_{ij,\,a}^{(1)} = 16/(\mu_i+\mu_j)^3$ at $\delta = 0$.
Its plot resembles that in Fig.\ \ref{sixplots}D.

The second integral is much simpler,
\Bea
	K_{ij,\,b}^{(1)} &=& \frac{1}{4\pi} \int d^3 r \;\left\{\; 
		e^{-\mu_i \,r^2_+/2}\; \, r_-^2 \; e^{-\mu_j\,r_-^2/2} +
		e^{-\mu_i \,r^2_-/2}\; \, r_+^2 \; e^{-\mu_j\,r_+^2/2} \right\}
			\nonumber \\
		&=& \ \left[ \frac{2\pi}{(\mu_i + \mu_j)^7} \right]^{1/2} 
			\left[\; 3\,(\mu_i + \mu_j) + 4 \,\mu_j^2\;\delta^2 \;\right]
			e^{-2\,\mu_i\mu_j \,\delta^2/(\mu_i + \mu_j)} \ ,
\Eea
which is also non-symmetric, but only because of the term proportional to $\delta^2$.
As a function of $\delta$ it looks like Fig.\ \ref{sixplots}E.

\pagebreak 
Almost as complicated as $K_{ij,\,a}^{(1)}$, the third integral is 
\Bea
	& & K_{ij,\,c}^{(1)} = \frac{1}{4\pi} \int d^3 r \;\left\{\; 
			e^{-\mu_j \,r^2_+/2}\; \, r_\pm \,({\bf r}_+ \cdot {\bf r}_-) \; e^{-\mu_i\,r_-^2/2} 
				\right. \nonumber \\
		& & \qquad\qquad\qquad\qquad\qquad\qquad +\ \left.
				e^{-\mu_j \,r^2_-/2}\; \, r_\pm \,({\bf r}_+ \cdot {\bf r}_-) \; e^{-\mu_i\,r_+^2/2} \right\}
				\nonumber \\
		& & \qquad = \ \frac{1}{\mu_i \mu_j(\mu_i+\mu_j)^4} \left\{\;
				2 \mu_j \; [\;(\mu_i+\mu_j)(4 \mu_i  - \mu_j) - 4 \mu_i^2 \mu_j \;\delta ^2 \;] 
				\;e^{-2\mu_i\;\delta ^2} \right. \nonumber \\
		& & \qquad\qquad\qquad\qquad\qquad\quad + \ 2 \mu_i \;
				[\;(\mu_i+\mu_j)(4 \mu_j  - \mu_i) - 4 \mu_i \mu_j^2 \;\delta ^2 \; ]
		 		\;e^{-2  \mu_j\; \delta ^2} \nonumber \\
		& & \qquad\qquad\qquad\qquad\qquad\quad \left. + \ 
				[\;(\mu_i+\mu_j)(\mu_i^2 - 8\,\mu_i\mu_j + \mu_j^2) + 8\,\mu_i^2\mu_j^2\;\delta ^2\;] 
		 		\;e^{-\frac{1}{2} (\mu_i+\mu_j)\; \delta ^2} \;\right\} \nonumber \\
		& &  \nonumber \\  
		& & \qquad\qquad + \ \frac{1}{2\,\delta\,\mu_i^2\,\mu_j^2} 
				\left[\frac{\pi}{2(\mu_i+\mu_j)^9}\right]^{1/2}	\times \\  
		& &  \nonumber \\  
		& & \qquad\qquad\qquad \left\{\;
				 2\,\mu_j^2 [\;(\mu_i+\mu_j)^2(4\mu_i + \mu_j) 
				+ 8\,\mu_i^2\,(\mu_i+\mu_j)(2\mu_i - \mu_j)\;\delta^2 - 16\,\mu_i^4\mu_j\; \delta^4\;]
				\right. \nonumber \\
		& & \qquad\qquad\qquad\qquad\qquad
				\times \;e^{-2\mu_i\mu_j\;\delta ^2/(\mu_i+\mu_j)} \;
				\text{Erf}\left( \sqrt{\frac{2}{\mu_i+\mu_j}}\;\mu_i\;\delta\;\right) \nonumber \\
		& & \qquad\qquad\qquad - \ 
				2\,\mu_i^2[\;(\mu_i+\mu_j)^2(4\mu_j + \mu_i) 
				+ 8\,\mu_j^2\,(\mu_i+\mu_j)(2\mu_j - \mu_i)\;\delta^2 - 16\,\mu_i\mu_j^4\; \delta^4\;]
				\nonumber \\
		& & \qquad\qquad\qquad\qquad\qquad 
				\times \;e^{-2\mu_i\mu_j\;\delta ^2/(\mu_i+\mu_j)} \;
				\text{Erfc}\left( \sqrt{\frac{2}{\mu_i+\mu_j}}\;\mu_j\;\delta\;\right) \nonumber \\
		& & \qquad\qquad\qquad  + \ 
				[\;(\mu_i+\mu_j)^2(\mu_i^2 + 5\,\mu_i\mu_j + \mu_j^2)\; 
					- 24\,\mu_i^2\mu_j^2(\mu_i+\mu_j)\;\delta^2 + 16\,\mu_i^3\mu_j^3\;\delta^4\;]
				 \nonumber \\
		& &	\qquad\qquad\qquad\qquad\qquad  
				\times (\mu_i-\mu_j) \;e^{-2\mu_i\mu_j\;\delta ^2/(\mu_i+\mu_j)} \;
				\;\left[1 + \text{Erf}\left( \frac{(\mu_i-\mu_j)} {\sqrt{2(\mu_i+\mu_j)}} \;\delta \;\right) 		
				\right] \nonumber \\
		& & \qquad\qquad\qquad  + \  \left. (\mu_i+\mu_j)^3\;[\,(\mu_i^2 + 3\,\mu_i\mu_j + \mu_j^2 )
				- 2\,\mu_i\mu_j(\mu_i+\mu_j)\;\delta^2\,] \right. \nonumber \\
		& &	\qquad\qquad\qquad\qquad\qquad  \times	\left. \;
				\text{Erfc}\left( \sqrt{(\mu_i+\mu_j)/2}\;\delta\;\right) \right\}  \nonumber \ ,
\Eea
which is surprisingly both symmetric, $K_{ji,\,c}^{(1)} = K_{ij,\,c}^{(1)}$, and non-singular:
$K_{ij,\,c}^{(1)} = 16/(\mu_i+\mu_j)^3$ at $\delta = 0$.
This integral as a function of $\delta$ looks like Fig.\ \ref{sixplots}B.

The fourth integral is also simple,
\Bea
	K_{ij,\,d}^{(1)} &=& \frac{1}{4\pi} \int d^3 r \;\left\{\; 
		e^{-\mu_j \,r^2_+/2}\; \, ({\bf r}_+ \cdot {\bf r}_-) \; e^{-\mu_i\,r_-^2/2} +
		e^{-\mu_j \,r^2_-/2}\; \, ({\bf r}_+ \cdot {\bf r}_-) \; e^{-\mu_i\,r_+^2/2} \right\}
			\nonumber \\
		&=&  \left[ \frac{2\pi}{(\mu_j + \mu_i)^7} \right]^{1/2}  \;
			[\;3 (\mu_j + \mu_i) - 4\mu_i\mu_j\;\delta^2 \,]
			\;e^{-2\mu_i\mu_j\;\delta ^2/(\mu_i+\mu_j)}  \ .
\Eea
Its $\delta$ dependence, Fig.\ \ref{sixplots}F, shows a relatively deeper minimum
than that depicted in Fig.\ \ref{sixplots}B.
The fifth and sixth integrals are already familiar,
\Bea
	K_{ij,\,e}^{(1)} &=& \frac{1}{4\pi} \int d^3 r \;\left\{\; 
		e^{-\mu_i \,r^2_+/2}\; \, r_\pm \; e^{-\mu_j\,r_-^2/2} +
		e^{-\mu_i \,r^2_-/2}\; \, r_\pm \; e^{-\mu_j\,r_+^2/2} \;\right\}
		\;=\; I_{ij,\,<r_\pm>}^{(1)} \\
	K_{ij,\,f}^{(1)} &=& \frac{1}{4\pi} \int d^3 r \;\left\{\; 
		e^{-\mu_i \,r^2_+/2}\; \; e^{-\mu_j\,r_-^2/2} +
		e^{-\mu_i \,r^2_-/2}\; \; e^{-\mu_j\,r_+^2/2} \;\right\}
		\;=\; I_{ij,\,<1>}^{(1)} \ .
\Eea
These last three integrals, $K_{ij,\,d}^{(1)}$ through $K_{ij,\,f}^{(1)}$,
are all symmetric in $i$ and $j$.


\begin{thebibliography}{99}

\bibitem{Carlson} Steven C. Pieper, R. B. Wiringa, and J. Carlson, Phys. Rev. C 70 
(2004) 054325-1. 

\bibitem{Bira} See, e.g., U. van Kolck, Prog. Part. Nucl. Phys. 43 (1999) 337.

\bibitem{EMC} European Muon Collaboration (J. Ashman et al.),
Z.Phys.C57 (1993) 211 and earlier papers cited there.

\bibitem{GMSS} T.\ Goldman, K.\ R.\ Maltman, G.\ J.\ Stephenson, Jr.,  and K.\ E.\ 
Schmidt, Nucl.\ Phys.\ A481 (1988) 621.

\bibitem{GBS} C.\ J.\ Benesh, T.\ Goldman, and G.\ J.\ Stephenson, Jr.,  Phys.\ 
Rev.\ C48 (1993) 1379  and Phys.\ Rev.\ C68 (2003) 045208.

\bibitem{BG} C.\ J.\ Benesh and T.\ Goldman, Phys.\ Rev.\ C55, (1997) 441.

\bibitem{Friar} See for example, J.\ L.\ Friar,  B.\ F.\ Gibson and G.\ L.\ Payne, 
Phys.\ Rev.\ C30 (1984) 1084.

\bibitem{piqk} E.g., T.\ Goldman and R.\ R.\  Silbar, Phys.\ Rev.\ C77 (2008) 065203.

\bibitem{fanwang} Fan Wang, Guang-han Wu, Li-jian Teng, and T.\ Goldman,
 Phys.\ Rev.\ Lett.\ 69 (1992) 2901.

\bibitem{eichten} E.\ Eichten, K.\ Gottfried, T. Kinoshita, K.\ D.\ Lane and T. -M.\ Yan, 
Phys.\ Rev.\ D17 (1978) 3090.

\bibitem{others} Some recent examples include: Takayuki Matsuki and Koichi Seo, 
Phys.\ Rev.\ D85 (2012) 014036; Stanislaw D.\ Glazek, ``Hypothesis of quark 
binding by condensation of gluons in hadrons'',  presented at LIGHTCONE 2011, 23 - 
27 May, 2011, Dallas, TX, arXiv:1110.1430; Jin-Hee Yoon, Byeong-Noh Kim, Horace 
W.\ Crater and Cheuk-Yin Wong, ``On the Mass Difference between pion and rho meson 
using a Relativistic Two-Body Model'', Talk presented at the Fifth Asia-Pacific Conference 
on Few-Body Problems in Physics, August 22-26, 2011, Seoul, Republic of Korea, to be 
published in Few-Body Systems, arXiv:1110.1598; M.\ Blank and A.\ Krassnigg, Phys.\ 
Rev.\ D84 (2011) 096014. 

\bibitem{nora} N.\ Brambilla, {\it et al.}, Eur.\ Phys.\ J.\ C71 (2011) 1534.

\bibitem{heller} L.\ Heller and J.\ A.\ Tjon, Phys.\ Rev.\ D35 (1987) 969; 
J.\ A.\ Carlson, L.\ Heller and J.\ A.\ Tjon, Phys.\ Rev.\ D37 (1988) 744.
 
\bibitem{PB} Carlos Pe\~{n}a and David Blaschke, Acta Phys.\ Pol.\ B, Proc.\ Suppl.\ 
5 (2012 963).

\bibitem{PGG} P.\ R.\ Page, T.\ Goldman, and J.\ N.\ Ginocchio, Phys.\ Rev.\ Lett.\  
86 (2001) 204.

\bibitem{Convolve} T.\ Goldman and R.\ R.\ Silbar, Phys.\ Rev.\ C85 (2012) 015203. 

\bibitem{HQET} N.\ Isgur and M.\ B.\ Wise, Phys.\ Lett.\ B237 (1990) 527,
Phys.\ Rev.\ D43 (1991) 819, Phys.\ Rev.\ Lett.\ 66 (1991) 1130, and
Nucl.\ Phys.\ B48 (1991) 276.

\bibitem{stringflip} See, e.g., M.\ Oka, Phys.\ Rev.\ D31 (1985) 2274,
Phys.\ Rev.\ D31 (1985) 2773, and subsequent related articles.

\bibitem{PDG} J.\ Beringer et al.\ (Particle Data Group), Phys.\ Rev.\ D86 (2012) 010001. 
A convenient access to this data is to go on-line to http://pdglive.lbl.gov/.

\bibitem{EJP} R.\ R.\ Silbar and T.\ Goldman, Eur.\ J.\ Phys.\ 32 (2011) 217. 




\end{thebibliography}
\end{document}